\documentclass[]{aa}  
\usepackage{ragged2e}

\usepackage{placeins}
\usepackage{graphicx}
\usepackage{tablefootnote}
\usepackage{hyperref}
\usepackage{academicons}
\usepackage{tikz,xcolor,hyperref}

\newcommand{\orcidauthorA}{0009-0009-8988-0537}
\newcommand{\orcidauthorB}{0000-0001-6264-140X}
\newcommand{\orcidauthorC}{0000-0003-0079-1239}
\newcommand{\orcidauthorD}{0000-0002-0273-218X}

\usepackage{orcidlink}
\usepackage{soul}

\hypersetup{
    colorlinks=true,
    citecolor=blue,
    linkcolor=blue,
    urlcolor=blue,
}
\usepackage{xcolor}
\usepackage{natbib}
\usepackage{txfonts}
\usepackage{comment}

\def\medd{$\dot{m}_{\rm Edd}$}
\def\mbh{$M_{\rm BH}$}
\def\psif{$\Psi_{\rm \lambda}(t)$} 
\def\gammaf{$\Gamma_{\rm \lambda}(\nu)$}
\def\swift{\textit{Swift}}

\def\sxvobs{$\sigma^2_{\rm XVobs}$}
\def\meansxv{$\overline{\sigma^2}_{\rm XVobs}$}

\def\snxvobs{$\sigma^2_{\rm NXVobs}$}
\def\ledd{$L_{\rm Edd}$}

\def\spin{$\alpha^{*}$}
\def\rg{$R_{\rm g}$}

\begin{document} 

\title{X-ray reverberation as an explanation for UV/optical variability in nearby Seyferts}

\author{M. Papoutsis \inst{1,2}
\and
I. E. Papadakis\inst{1,2,3}
\and
C. Panagiotou\inst{4} 
\and 
M. Dovčiak\inst{5}
\and
E. Kammoun\inst{6, 7}
}

\titlerunning{UV and optical  AGN variability}
\authorrunning{M. Papoutsis et al.}

\institute{Department of Physics, University of Crete, 71003 Heraklion, Greece \email{mpapoutsis@physics.uoc.gr}
\and 
Institute of Astrophysics, FORTH, GR-71110 Heraklion, Greece
\and
Department of Physics and Institute of Theoretical and Computational Physics, University of Crete, 71003 Heraklion, Greece
\and
MIT Kavli Institute for Astrophysics and Space Research, Massachusetts Institute of Technology, Cambridge, MA 02139, USA
\and
Astronomical Institute of the Academy of Sciences, Boční II 1401, CZ-14100 Prague, Czech Republic
\and
IRAP, Université de Toulouse, CNRS, UPS, CNES, 9, Avenue du Colonel Roche, BP 44346, F-31028, Toulouse Cedex 4, France
\and
INAF—Osservatorio Astrofisico di Arcetri, Largo Enrico Fermi 5, I-50125 Firenze, Italy
}


   \date{Received 14 November 2023; accepted 30 August 2024}

 
  \abstract
   {Active galactic nuclei (AGNs) are known to be variable across all wavelengths. Significant observational efforts have been invested in the last decade in studying their ultraviolet (UV)  and optical variability. Long and densely sampled, multi-wavelength monitoring campaigns of numerous Seyfert galaxies have been conducted with the aim of determining the X-ray/UV/optical continuum time lags. Time-lag studies can be used to constrain theoretical models. The observed time lags can be explained by thermal reprocessing of the X-rays illuminating the accretion disc (known as the X-ray reverberation model). However, the observed light curves contain more information that can be used to further constrain physical models.}
   {Our primary objective is to investigate whether, in addition to time lags, the X-ray reverberation model can also explain the UV/optical variability amplitude of nearby Seyferts.}
   {We measured the excess variance of four sources (namely Mrk\,509, NGC\,4151, NGC\,2617, and Mrk\,142) as a function of wavelength using data from archival long, multi-wavelength campaigns with \swift\, and ground-based telescopes. We also computed the model excess variance in the case of the X-ray reverberation model by determining the disc's transfer function and assuming a bending power law for the X-ray power spectrum. We tested the validity of the model by comparing the measured and model variances for a range of accretion rates and X-ray source heights.}
   {Our main result is that the X-ray thermal reverberation model can fit both the continuum, UV/optical time lags, as well as the variance (i.e. the variability amplitude) in these AGNs, for the same physical parameters. Our results suggest that the accretion disc is constant and that all the observed UV/optical variations, on timescales of days and up to a few weeks, can be fully explained by the variable X-rays as they illuminate the accretion disc.}
   {}
   

   \maketitle
%
\section{Introduction}

Over the past ten years, enormous observational efforts have been invested in the multi-wavelength monitoring of many nearby Seyferts with the \textit{Neil Gehrels Swift} Observatory (hereafter, \swift). A large number of observational campaigns have already been conducted, monitoring the continuum variations from X-ray to ultraviolet (UV) and optical, at high cadence and over long periods of time \cite[e.g.] []{Shappee14,McHardy14,McHardy18,Fausnaugh16,Cackett2020,Hernandez-santisteban20,Kara21}. The main objective of these observations is to estimate the cross-correlation between the variations seen in the X-ray and in the UV/optical and to constrain models for the continuum emission in active galactic nuclei (AGNs).

The hypothesis of X-ray illumination of the accretion disc was proposed many years ago to explain the detection of the Fe K$\alpha$ emission line around 6.4~keV and the spectral hardening above $\sim 10$ keV in the Ginga spectra of AGNs \cite[ e.g.][]{Pounds90,George91}. It was immediately recognised that if the accretion disc is illuminated by X-rays and the X-rays are variable, then the X-ray and disc emission should be variable in a correlated way -- but with a delay that should be representative of the light travel time from the X-ray corona to the disc. This was indeed observed in the first multi-wavelength monitoring campaign of NGC\,5548, which showed that the UV and X-ray flux variations were positively correlated and simultaneous to within a few days \citep{Clavel92}. 

We  carried out a detailed study of the X-ray reverberation of the accretion disc in AGNs over the last few years. \cite{Kammoun19} and \cite{Kammoun211} (hereafter, K21a) studied the disc response to illuminating X-rays, taking into account all relativistic as well as the disc ionisation effects and then fitting (successfully) the time-lags as a function of wavelength in many sources \citep{Kammoun212, Kammoun23}. \cite{Dovciak22} presented {\tt KYNSED}, a new spectral model that can be used to fit the broadband spectral energy distributions (SEDs) of AGNs, from optical to X-rays, in the case where X-rays are illuminating the disc. These authors used the model to  successfully fit the average SED of NGC\,5548, using data from the STORM campaign \citep{Fausnaugh16}. \cite{Papadakis22} also used {\tt KYNSED} and showed that X-ray illumination of the accretion disc can explain the discrepancy between the theoretical and the observed accretion disc sizes in micro-lensed quasars. Moreover, \cite{K24} employed {\tt KYNSED} and showed that X-ray illumination can explain both the time-averaged and time-resolved SEDs of NGC\,5548. 

Although the major objective of the monitoring campaigns so far has been the study of the delays between the optical/UV variations and those in the X-rays, the resulting light curves hold a wealth of additional information that can be used to constrain theoretical models. The time lags depend mainly on the width of the disc response function to a burst of X-ray radiation, but not on its amplitude. On the other hand, the variability amplitude in the optical/UV bands should depend on the amplitude of the disc response function and is an additional observational quantity that models should be able to explain. In fact, a model should be able to explain both the time lags and the variability amplitude of the observed light curves at the same time. This is was the approach adopted by \cite{Panagiotou20,Panagiotou22}  with respect to the STORM data of NGC\,5548. 

Our work is similar to \citet[P22 hereafter]{Panagiotou22}, with the main difference being that P22 studied power spectra, whereas we focus on the variance of the light curves. In particular, we chose four sources from the sample of \citet[K21b hereafter]{Kammoun212} and computed the excess variance of the light curves in various bands. Since the variance is the integral of the power spectrum, we used the P22  model transfer functions to compute the model variance in each band, assuming the model parameters that K21b determined by fitting the time-lags of the same sources. By comparing the model with the observed variances, we investigate whether the X-ray reverberation model can fit both the optical/UV time lags and observed variance in AGNs.

This work is organised as follows. In Section \ref{sec:sample} ,we present the sample used in our work. In Section \ref{sec:rev_model}, we describe the X-ray reverberation model. In Sections \ref{sec:observed} and \ref{sec:model}, we explain how we compute the observed and model excess variance respectively. Finally, we compare the model variance spectra with the observations in Sect. \ref{sec:comparison}. We summarise our work and discuss our results in Sect. \ref{sec:discussion}.

\section{The sample}
\label{sec:sample}
The sources we study are taken from K21b and are listed in Table\,\ref{table:sourcestable}. We omitted NGC\,5548, which was already studied by P22, NGC\,4593, and NGC\,7469 because the light curves are not available online. The luminosity distance, $ D_{\rm L}$, the black hole mass, $M_{\rm BH}$, the 2-10 keV X-ray luminosity, $L_{\rm Xobs, Edd}$, and photon index, $\Gamma$, are taken from K21b. The bolometric luminosity is used to compute the bending frequency in the X-ray power spectrum (see Sect.\,\ref{sec:psdx}). It is computed assuming that $L_{\rm Bol}=\dot{m}_{\rm Edd}L_{\rm Edd}$, where \ledd\ is the Eddington luminosity and \medd\ is the accretion rate in Eddington units. We assumed the accretion rate estimate of \cite{Edelson2019} for Mrk\,509, the values listed in Table\,1 of K21b for NGC\,4151 and NGC\,2617, and 0.37 for Mrk\,142 (with their computation explained in Sect.\,\ref{sec:mrk142}). These estimates are based on the use of the luminosity at a particular wavelength and various bolometric correction factors to transform it to $L_{\rm Bol}$.  

\begin{center}
   \begin{table}
      \caption[]{Redshift, $z$, luminosity distance, $D_{\rm L}$, BH mass, $M_{\rm BH}$, 2-10 keV X-ray luminosity, $L_{\rm Xobs, Edd}$, and photon index, $\Gamma$, of the sources in our sample.}
         \label{table:sourcestable}    
\[\arraycolsep=5pt\def\arraystretch{1.2}
         \begin{array}{lllllll}
            \hline
            \noalign{\smallskip}
            \textrm{Source}  & z^a &   D_{\rm L}^b & M_{\rm BH}^c &  L^d_{\rm Bol} & L^e_{\rm Xobs, Edd} & \Gamma \\
            \noalign{\smallskip}
            \hline
            \noalign{\smallskip}
            \textrm{Mrk }509 & 0.034  & 151 & 1.12 & 7.1 & 8.3 & 1.75 \\
            \textrm{NGC }4151 & 0.003  & 17 & 0.36 & 0.9 & 1.7 &  1.4  \\
            \textrm{NGC }2617& 0.014  & 62 & 0.32  & 0.4 & 2.6 & 1.8 \\
            \textrm{Mrk }142& 0.045  & 199 & 0.04 & 1.85  & 12.4 & 2.4  \\
            \noalign{\smallskip}
            \hline            
         \end{array}
\]
\tablefoot{($^a$) taken from NASA/NED\tablefootnote{The NASA/IPAC Extragalactic Database (NED) is funded by the National Aeronautics and Space Administration and operated by the California Institute of Technology.}, ($^b$) in Mpc, ($^c$) in units of 10$^8$M$_{\odot}$, ($^d$) in units of 10$^{44}$ (erg/s), ($^e$) in units of $10^{-3} L_{\rm Edd}$. $D_{\rm L}$,  $M_{\rm BH}$, $L_{\rm Bol}$, $L_{\rm Xobs, Edd}$, and $\Gamma$ are taken from K21b for Mrk\,509, NGC\,4151, and NGC\,2617, while those for Mrk\,142 are computed as explained in Sect. \ref{sec:mrk142}.}
\end{table}
\end{center}

\section{The X-ray reverberation model}
\label{sec:rev_model}
We assumed a point-like X-ray source at height, $h$, on the rotational axis of the central black hole (BH). The source emits isotropically (in its rest frame) a spectrum of the form $F_{\rm X}(t)=N(t) E^{-\Gamma} \exp\left(-E/E_{\rm cut}\right)$,  where $\Gamma$ and $E_{\rm cut}$ are assumed to be constant. X-rays illuminate the accretion disc, which co-rotates with the BH. A fraction of the X-rays falling on the disc will be re-emitted as X-rays (this is the 'disc reflection component'), while the rest will be absorbed by the disc. They will thermalise in the disc and act as an extra source of heating. Consequently, the local disc temperature and emission will increase. As long as the X-rays are variable (which they indeed are in AGNs), the disc flux in the UV/optical wavebands will also be variable. 

Following K21a, the disc flux in a wavelength $\lambda$, in the case of the X-ray illuminating discs, will be given by:
\begin{equation}
    F_{\rm \lambda}(t) = F_{\rm \lambda, NT} +
     \int_{0}^{\infty} \Psi_{\lambda}(t')L_{\rm Xobs,Edd}(t-t')dt' \;,
\label{eq:discflux}
\end{equation}

\noindent where $F_{\rm \lambda}$ is the total disc flux at wavelength $\lambda$, $F_{\rm \lambda, NT}$ is the flux of a standard accretion disc when there is no illumination \cite[][]{NT73}, and $L_{\rm Xobs,Edd}$ is the observed X-ray luminosity in the 2--10 keV band, in units of Eddington luminosity (\ledd). $\Psi_{\lambda}(t)$ is the disc response to an X-ray flash, defined by Eq. (3) in K21a. It is equal to the flux that the disc emits due to X-ray heating at time $t$ after the X-ray flash. Therefore,  the convolution integral in the right part of Eq.\,(\ref{eq:discflux}) is equal to the variable, X-ray thermally reprocessed flux of the disc when the X-rays are constantly variable.

\subsection{The UV/optical power spectrum of X-ray illuminated discs}

Equation\,(\ref{eq:discflux}) is significantly simplified when expressed in the frequency domain. It can be shown that \citep[e.g.][P22]{Papadakis16},
\begin{equation}
    \textrm{PSD}_{\rm \lambda}(\nu) = |\Gamma_{\rm \lambda}(\nu)|^{2} \cdot \textrm{PSD}_{\rm X,Edd}(\nu) \; ,
\label{eq:psds}
\end{equation}

\noindent where $\textrm{PSD}_{\lambda}(\nu)$ is the power spectrum at wavelength $\lambda$, $\textrm{PSD}_{\rm X,Edd}$ is the X-ray power spectrum when the light curve is in units of \ledd, and $\Gamma_{\lambda}(\nu)$ is the transfer function, which is equal to the Fourier transform of the response function, \psif. Just as \psif, we know that \gammaf\, depends on the geometric and physical properties of the X-ray source and the accretion disc; for example, \mbh, \medd, X-ray luminosity, X-ray source height, and more (see P22).
Consequently, by fitting the UV/optical PSDs when the X-ray PSD is known, we can determine \gammaf\, and, hence, the physical parameters of the central source in AGNs. 

This was the approach taken in P22 in their study of the optical and UV PSDs of NGC\,5548.  However, the estimation of the power spectrum is a non-trivial task, mainly because the light curves are not evenly sampled. Fitting the observed PSDs with Eq.\,(\ref{eq:psds}) is not easy either, as shown in P22. We plan to perform a full PSD analysis of many sources in the near future (Panagiotou et al, in prep.). In the present work, we follow a simpler approach, namely: we measure the variance of the sources in the sample in the optical/UV bands and we fit the resulting `variability spectrum' (i.e. variance versus photon frequency), assuming a wide range of model parameters. Then we compare our results with the best-fit results of K21b to investigate whether X-ray reverberation can explain both the time lags as well as the variability amplitude in AGNs with the same physical parameters. 

\subsection{The variance  of X-ray illuminated discs in the UV/optical bands} 
\label{sec:varmodel}

The power spectrum of a random process, PSD($\nu)$, is defined in such a way that
\begin{equation}
    \sigma^{2} = \int_{0}^{\infty} \textrm{PSD}(\nu) d\nu, 
\label{s2definition}
\end{equation}

\noindent 
where $\sigma^2$ is the variance of the process. The variance we measure using observed light curves is not equal to the integral of the power spectrum over all frequencies, as we expect there to be intrinsic variations on long (and short) timescales that are not fully sampled by the observations. \cite{Allevato13} showed that the variance (at a spectral band $\lambda$) measured using a light curve of duration $T$ and bin size of $\Delta t$ is an estimator of the so-called band variance, defined as:
\begin{equation}
    \sigma^{2}_{\lambda, \textrm{band}} = \int_{1/T_{\rm max}}^{1/T_{\rm min}}  \textrm{PSD}_{\lambda}(\nu) d\nu \; ,
\label{eq:s2_band_1}
\end{equation}
where  $T_{\rm max}=T$ and $T_{\rm min}=2\times \Delta t$, the longest and shortest sampled variability timescales respectively. 
Using Eq.\,(\ref{eq:psds}), the equation above can be rewritten as follows
\begin{equation}
    \sigma^{2}_{\lambda, \textrm{band}} = \int_{1/T_{\rm max}}^{1/T_{\rm min}} |\Gamma_{\lambda}(\nu)|^{2} \cdot \textrm{PSD}_{\rm X,Edd}(\nu) d\nu. 
\label{eq:s2_band_2}
\end{equation}
This equation can be used to compute the model variance of the UV/optical light curves. The comparison between the model and the observed variance can test the validity of the model and can provide the set of model parameters that fit the data (i.e. the light curve variances) best.  

\begin{center}
\begin{table*}
      \caption[]{Variance measurements, i.e.\, $\overline{ \sigma^{2}}_{\rm HX,NXVobs}$, for the hard X-ray band (HX), and \meansxv\, in the UV/optical bands, in units of $10^{-3}$ (mJy)$^2$.}
         \label{table:variancetable}    
\[\arraycolsep=6pt\def\arraystretch{1.0}
         \begin{array}{lllll}
            \hline
            \noalign{\smallskip}
            \textrm{Band} & \textrm{Mrk 509} & \textrm{NGC 4151} & \textrm{NGC 2617} &  \textrm{Mrk 142}\\
             &   \textrm{(E19)} &  \textrm{(E17)} &  \textrm{(F18)} &  \textrm{(C20)}\\
            \noalign{\smallskip}
            \hline
            \noalign{\smallskip}
            \textrm{HX} & 0.015\pm 0.003 (20,12,1.03)  & 0.027\pm 0.004 (21,3,0.21)  &  0.16 \pm 0.04 (11,13,1.47) &  0.19 \pm 0.02 (19,5.5,0.6) \\
            \textrm{W2}/1928 \textrm{\AA}  &  38 \pm 11 (20,12,1.12) &  7 \pm 2 (21,3,0.27) &  96 \pm 42 (10,13,1.47)  &  1.7 \pm 0.4 (17,6,0.74) \\
            \textrm{M2}/2236\textrm{\AA}  & 64 \pm 26 (20,12,1.2)  & 8 \pm 3 (21,3,0.27) & 80 \pm 34 (11,13,1.52)  & 0.8 \pm 0.3 (16,6,0.74)\\
            \textrm{W1}/2600\textrm{\AA}  &  29 \pm 17 (20,12,1.17)  & 10 \pm 4 (21,3,0.25) & 97 \pm 44 (10,13,1.5) &  1.3 \pm  0.4 (17,6,0.74)\\
            \textrm{SU}/3467\textrm{\AA} &  57 \pm 15  (20,12,1.07) & 33 \pm 16  (21,3,0.25) & 185 \pm 84 (10,13,1.51) & - \\
            \textrm{u}/3540\textrm{\AA}  &  -  & - &  - & 0.5  \pm  0.2 (30,6,0.68)\\
            \textrm{SB}/4392\textrm{\AA}  &   35 \pm 10  (20,12,1.06) & 55 \pm 16 (21,3,0.22) &  67 \pm 43 (10,13,1.52) &  1.1 \pm 0.5 (17,6,0.66) \\
            \textrm{g}/4770\textrm{\AA}  &  - & - & - & 0.5 \pm 0.1  (29,6,0.57)\\
            \textrm{SV}/5468\textrm{\AA}  &  17 \pm 9 (20,12,1.11)  & 90 \pm 150 (5,12,0.22) & 32 \pm 32 (10,13,1.42) & -  \\            
            \textrm{r}/6215\textrm{\AA}  & - & - & 9 \pm 3 (14,6,0.76) & 0.3 \pm 0.1 (31,6,0.56)\\
            \textrm{i}/7545\textrm{\AA}  &  - & -  & 10 \pm 4  (15,6,0.7) & 0.4 \pm 0.1 (30,6,0.57)\\
            \textrm{z}/8982\textrm{\AA}  &  - &  -  & 19 \pm 7  (14,6,0.73) & - \\
            \noalign{\smallskip}
            \hline
         \end{array}
\]
\tablefoot{Numbers in the parentheses next to the variance measurements list the number of segments, their duration, $T_{\rm max}$ (days), and the average $\Delta t$ (days) of the light curves. SU, SB, and SV denote the respective \swift\ bands. Light curves are taken from: \cite{Edelson2019}, \cite{Edelson_2017}, \cite{Fausnaugh_2018}, and \cite{Cackett2020} (E19, E17, F18, and C20, respectively). The number next to each band name indicates the effective wavelength, taken from C20, except for the $z$ band which is taken from F18. }
\end{table*}
\end{center}

\section{Observed variance}
\label{sec:observed}
\subsection{The excess variance}

One of the most common estimators of the intrinsic variance is the excess variance, $\sigma^{2}_{\rm XVobs}$, 
\begin{equation}
    \sigma^{2}_{\rm XVobs} = \frac{1}{N}
\sum_{i=1}^{N}[(x_{i}-\bar{x})^2 - \sigma_{{\rm err}, i}^{2}],
\label{eq:sxv}
\end{equation}

\noindent according to \cite{Allevato13}, where $x_{i}$, $\bar{x}$, and $\sigma_{{\rm err}, i}$ are the count rate, the sample mean, and the error bar of the light curve points, respectively.  
We could compute the excess variance using the full light curve, however, single \sxvobs\ measurements are not useful. Their distribution is very asymmetric and very broad, and their variance is unknown. For that reason, we divided the light curves into many segments, computing \sxvobs\ for each one of them and then calculated their mean, \meansxv, and its error using Eqs. (12) and (13) in \cite{Allevato13}. According to these authors, as long as the signal-to-noise ratio of the light curves is higher than 3 (which is the case for all light curves, in all bands, in our case), the mean variance should be approximately Gaussian distributed, and its intrinsic standard deviation (i.e. its error) should be equal to the error of the mean variance, as long as the number of segments is $\gtrsim 20$. 

Figures\,\ref{fig:mrk509_lightcurve}, \ref{fig:ngc4151_lightcurve}, \ref{fig:ngc2617_lightcurve}, and \ref{fig:mrk142_lightcurve} show the light curves in X-rays and in at least one UV/optical band for all sources. We indicate the light curve segments we used to compute the variance, and we also plot \snxvobs\ of each segment. The \meansxv\ measurements and their error
are listed in Table\,\ref{table:variancetable}. The  first number in the parenthesis next to \meansxv\ indicates the number of light curve segments we considered, 
while the second and third numbers list the duration of the segments, $T_{\rm max}$, and the average $\Delta t$ of the light curves (note that, despite the excellent cadence of the \swift\ observations, the resulting light curves are not evenly sampled). The number of segments in some bands is smaller than 20. The shorter the $T_{\rm max}$, the larger the number of segments would be. However, if $T_{\rm max}$ is too short, then the source variance is smaller than the Poisson noise variance and the \sxvobs\ error bar is large. Therefore, the $T_{\rm max}$ values we chose were a compromise between the need for the largest possible number of segments and the detection of the intrinsic variance with the smallest possible uncertainty. 

\subsection{Bias in the excess variance}
\label{sec:s2bias}

In the case of red-noise-like PSDs (as is the case with AGNs), variations with timescales longer than $T_{\rm max}$ can contribute to the measured variance, even if they are not fully sampled (this is known as the red-noise leakage). Consequently, the sample variance, $\sigma^2_{\rm XVobs}$, may overestimate the intrinsic variance, $\sigma^{2}_{\rm \lambda, band}$. This introduces a bias that we need to take into account when comparing the model band variance (computed using Eq.\,\ref{eq:s2_band_2}) with \meansxv\ (i.e. the mean of all variances we compute for each segment using Eq.\,{\ref{eq:sxv}). According to \cite{Allevato13} the bias depends on the PSD slope (say $\beta$) in the frequency range between $1/(2.1T_{\rm max})$ and $1/T_{\rm max}$, in such a way that, 

\begin{equation}
    \sigma^{2}_{\rm \lambda, band} = \overline{\sigma^{2}}_{\rm XVobs} \times 0.48^{(\beta - 1)}.
     \label{eq:bias}
 \end{equation}

\noindent This means that before comparing the variance models with observations we need to take into account the correction factor of $0.48^{(\beta - 1)}$. In all cases where it is necessary to correct for the bias of the measured variance, we can compute $\beta$ by measuring the slope of a straight line between the model PSD points \{log($1/(2.1T_{\rm max})$), log[PSD$(1/(2.1T_{\rm max}))]\}$ and \{log($1/T_{\rm max})$, log[PSD$(1/T_{\rm max})]\}$. A straight line can fit well the model PSDs in this relatively small frequency range in all cases we consider here. We note that $\beta \sim 1 - 3$ for the sources studied here, with larger $\beta$ values corresponding to larger wavelengths.

\section{Model variance}
\label{sec:model}
According to Eq.\,(\ref{eq:s2_band_2}) we need to compute the disc transfer function, $\Gamma_{\lambda}(\nu)$, and to know the X-ray power spectrum in order to compute the model variance that we will then compare with the observations. We describe below how we compute the model variance. 

\subsection{The disc transfer function}
\label{sec:transfer}

P22 studied in detail the accretion disc transfer functions in the case of X-ray-illuminated discs and showed that they can be well approximated by the following analytical formula:
\begin{equation}
    |\Gamma_{\lambda}(\nu)|^{2} = N \frac{1}{1+
   (\frac{\nu}{\nu_{\rm b}})^s} \; ,
\label{eq:gamma}
\end{equation}
where the normalisation $N$ is in units of (erg/s/cm$^2$/$\AA$)$^2$. According to the equation above, $|\Gamma_{\lambda}(\nu)|^{2}$ is constant at low frequencies and then it bends to a power-law shape, with a slope of $s$ at frequencies higher than a characteristic frequency, $\nu_b$. Equations\,(\ref{eq:s2_band_2})  and (\ref{eq:gamma}) show that \gammaf\ works like a low-pass frequency filter: all X-ray variations on time scales longer than $\sim 1/\nu_b$ will also appear in the optical/UV bands. Shorter timescale X-ray variations are smeared out due to the X-ray reprocessing process. Consequently, the optical/UV PSD is suppressed at frequencies higher than $\sim \nu_b$. The parameters $N$, $s,$ and $\nu_{\rm b}$, are functions of the wavelength and the system's physical parameters: $M_{\rm BH}$, $\dot{m}_{\rm Edd}$; the height and luminosity the X-ray source, $h$ and $L_{\rm Xobs,Edd}$; and the spectral photon index, $\Gamma$; the disc inclination $\theta$; and the BH spin $\alpha^{*}$ (see Appendix A of P22 for the exact forms of the analytical functions). 

Equation\,(\ref{eq:gamma}) is a good approximation of the transfer function at low frequencies. 
It works well for the range of physical parameters and wave bands considered here,  at frequencies below $\sim 4.3$ day$^{-1}$ (see Sect. 2.4 in P22), which corresponds to a sampling rate of around 0.23 day. This is comparable/smaller than the sampling rate of the sources in our sample (see Table\,\ref{table:variancetable}).

\subsection{The X-ray power spectrum}
\label{sec:psdx}

We assume that the X-ray PSD has a bending power law shape, with a slope of $-1$ at low frequencies:
\begin{equation}
    \textrm{PSD}_{\rm X}(\nu) = \frac{A}{\nu} \cdot \frac{1}{1+ (\frac{\nu}{\nu_{\rm b, X}})^{s_{\rm X}-1}} \;,
\label{eq:PSD_X}
\end{equation}
where $A$ is the normalisation, $\nu_{\rm b, X}$ the bending frequency, and $s_{\rm X}$ the spectral index above the bend. 
Equation\,(\ref{eq:PSD_X}) with $s_{X}=2$ fits  the X-ray power spectra of AGNs well \cite[e.g.][]{Uttley02,Markowitz03,McHardy04,Gonzalez-Martin2012}. We note that the use of a different slope, from $1.5$ up to $3$, does not  significantly affect our main results. 

To compute the bend frequency, $\nu_{\rm b, X}$, we assume the \cite{McHardy06} relation of, 
\begin{equation}
    \textrm{log}T_{\rm b,X} = a \times \textrm{log}(M_{\rm BH,6}) - b \times \textrm{log}(L_{\rm Bol,44}) + c\;,
\label{eq:logTb}
\end{equation}
where $a = 2.17^{+0.32}_{-0.25}$, $b = 0.9^{+0.3}_{-0.2}$, $c = -2.42^{+0.22}_{-0.25}$, $T_{\rm b, X}=1/\nu_{\rm b,X}$ (in days), $M_{\rm BH,6}$ is the black hole mass in units of $10^{6}M_\odot$, and $L_{\rm Bol,44}$ is the bolometric luminosity in units of $10^{44}$erg/s. 

As for the determination of the PSD normalisation, we can use the X-ray excess variance to compute $A$ in Eq.\,(\ref{eq:PSD_X}) since, according to Eq.\,(\ref{eq:s2_band_1}), 
\begin{equation}
    \sigma^{2}_{\rm HX,band} = \int_{1/T_{\rm max}}^{1/T_{\rm min}} \frac{A}{\nu} \cdot \frac{1}{1+ (\frac{\nu}{\nu_{\rm b, X}})^{s_{\rm X}-1}} d\nu \;.
\label{eq:s2_band_X}
\end{equation}

\noindent 
The row `HX'\footnote{HX refers to the `hard' X-ray band, which is not the same in all sources. E19 extracted the Mrk\,509 light curve in the 1.5--10 keV band, while we use the 5--10 keV band light curve of E17 for NGC\,4151 (the source is absorbed in lower energies). As for NGC\,2617 and Mrk\,142, we consider the 0.3--10 keV band light curves of F18 and C20. Note: the broadband variance in AGNs does not depend significantly on the energy limits of the band (see e.g. Fig.\,2 in \cite{Ponti12}).} in Table\,\ref{table:variancetable} lists the normalised excess variance of the X-ray light curves. Just like with the UV/optical light curves, we divided the X-ray light curves into many segments (their number indicated in parenthesis, next to the sample variances), and we computed the normalised excess variance of each one. The X-ray variance we list in Table\,\ref{table:variancetable} is the mean of the individual variance measurements of each light curve segment, $\overline{\sigma^{2}}_{\rm HX,NXVobs}$.


We can use the observed X-ray variance and the equation above to compute the PSD amplitude, $A$. However, as we mentioned in Sect.\,\ref{sec:s2bias},  $\sigma^{2}_{\rm HX,NXVobs}$ may be a biased estimator of $\sigma^{2}_{\rm HX,band}$, due to red noise leakage effects.  We therefore used Eq.\,(\ref{eq:bias}) and we corrected the observed variances for bias. Using then Eqs. (\ref{eq:s2_band_X}) (with $s_{\rm X}=2$) it is straightforward to show that
\begin{equation}
    A = \sigma^{2}_{\rm HX,NXVobs} \times 0.48^{(\beta -1)} \times \left[ \textrm{ln}\left(\frac{\nu_{\rm b, X} T_{\rm max} + 1}{\nu_{\rm b, X} T_{\rm min} + 1}\right) \right]^{-1},
\label{eq:PSD_X_norm}
\end{equation}
 
\noindent where $\beta$ is the slope of the X-ray PSD, which was computed as described in Sect.\,\ref{sec:s2bias}. Computed in this way, $A$ is the PSD normalisation for a light curve which is normalised to its mean. However, the PSD in Eq.\,(\ref{eq:s2_band_2}) should be the power spectrum of the X-ray light curve in units of \ledd. The PSD in this case can be easily calculated by multiplying $A$ by $L_{\rm Xobs,Edd}^2$. 

When combining all the above, the final equation that gives the X-ray power spectrum is as follows: 
\begin{equation}
{\rm PSD}_{\rm X,Edd}(\nu)=\frac{L_{\rm Xobs,Edd}^2\times \sigma^{2}_{\rm HX,NXVobs} \times 0.48^{(\beta-1)} }
{\ln(\frac{\nu_{\rm b, X} T_{\rm max} + 1}{\nu_{\rm b, X} T_{\rm min} + 1})}
\cdot
\frac{\nu^{-1}}{1+ (\frac{\nu}{\nu_{\rm b, X}})}.
\label{eq:PSD_XEdd}
\end{equation}

\subsection{Calculation of the model band variance}
\label{sec:modelvar}

We calculate the model band variance for the UV/optical bands, $\sigma^2_{\rm \lambda,band}$, using Eq.\,(\ref{eq:s2_band_2}). The transfer function and the X-ray power spectrum are calculated as described in Sects.\, \ref{sec:transfer} and \ref{sec:psdx}. However, there are a few more issues we must consider, before using Eq.\,(\ref{eq:s2_band_2}) to compute $\sigma^2_{\rm \lambda,band}$.

First, since the variance in the UV/optical bands is given in physical units, we have also to consider the effects of Galactic (and host galaxy) interstellar absorption, which affects mainly the UV bands. To account for the effects of the Galactic interstellar reddening, we must multiply $\sigma^{2}_{\rm \lambda, band}$ by the square of the absorption factor $ {f}_{\rm Gal,\lambda}=10^{ - A_{\lambda}/2.5}$, where $A_\lambda$ is the total extinction at a wavelength $\lambda$. Extinction was calculated using the reddening curves of \cite{Cardelli89}, with $R_{V}=3.1$. The Galactic $E(B-V)$ values for each source in our sample were taken from \cite{Schlafly_2011}. It is necessary to correct for reddening caused by the host galaxy in the case of NGC\,4151, as we explain in Sect.\,\ref{sec:comparison}. In this case, $\sigma^{2}_{\rm \lambda, band}$ should be multiplied by a second absorption factor as well, say $f^{2}_{\rm host,\lambda}$, which for simplicity was calculated like $f^{2}_{\rm Gal,\lambda}$ (for a different $E(B-V)$, of course).

Second, the response functions of K21a that P22 used to compute the disc transfer functions assumed a distance of 1Mpc. Therefore, $\sigma^{2}_{\rm \lambda, band}$ should be  divided by $D^{4}_{\rm L}$, where $D_{\rm L}$ is the luminosity distance (in Mpc).

Finally, to compare the model variance with \meansxv\, we need to correct for bias. We use Eqs.\,(\ref{eq:psds}), (\ref{eq:gamma}), and (\ref{eq:PSD_XEdd}) to calculate the model power spectrum in each band $\lambda$ and then we compute the slope $\beta$ as described in Sect.\ref{sec:s2bias}. We then divide the model variance with a  factor of 0.48$^{(\beta-1)}$ to account for the bias of the variance measurements (given the model).

Taking into account the discussion in previous sections and the points above, the X-ray reverberation model band variance in the optical/UV bands becomes:
\begin{equation}
    \sigma^{2}_{\lambda, \textrm{band}} = 
\frac{f^{2}_{\rm Gal,\lambda}\times f^{2}_{\rm host,\lambda}}{D^{4}_{\rm L} \times 0.48^{(\beta-1)}}
\int_{1/T_{\rm max}}^{1/T_{\rm min}} 
|\Gamma_{\lambda}(\nu)|^{2} \cdot \textrm{PSD}_{\rm X,Edd}(\nu) \cdot d\nu, 
\label{eq:s2_band_final}
\end{equation}

\noindent where $\Gamma_{\lambda}(\nu)$ and $\rm PSD_{ \rm X,Edd}(\nu)$ are given by Eqs.\,(\ref{eq:gamma}) and (\ref{eq:PSD_XEdd}), respectively. The model variance defined by the equation above is in units of (ergs/s/cm$^{2}$/\AA)$^2$. It is computed using the transfer functions of P22, which are given in these units. However, we chose to compute the variance of the light curves in units of (mJy)$^2$, because in this case, the variance versus photon frequency relation is flatter. This makes it easier to compare with the broadband spectral energy distribution of the objects. Therefore, in order to model the observed variance spectra, we need to transform the model $\sigma^{2}_{\lambda, \textrm{band}}$ as computed above, to model variance in units of (mJy)$^2$, $\sigma^{2}_{\nu, \textrm{band}}$. This is done in the same way one transforms flux measurements from one to the other unit.

\section{Fitting the model to the data}
\label{sec:comparison}

\subsection{Data:\  Observed variance spectra}
\label{sec:observedvariancespectra}

Black points in Figs.\, \ref{fig:mrk509}, \ref{fig:ngc4151}, \ref{fig:ngc2617}, and \ref{fig:mrk142} show the observed variance spectra (i.e. \meansxv, taken from Table\,\ref{table:variancetable}, plotted as a function of photon frequency), for Mrk\,509, NGC\,4151, NGC\,2617, and Mrk\,142, respectively. Points in red in Figs.\,  \ref{fig:ngc4151} and \ref{fig:ngc2617} indicate variance measurements computed using different duration segments (the second number in parentheses in Table\,\ref{table:variancetable}). The $x-$axis shows the rest frame frequency, that is, $\nu_{\rm rf} = \nu_{\rm obs} (1+z)$, where $\nu_{\rm obs}=c/\lambda_{\rm obs}$, with $\lambda_{\rm obs}$  values listed in Table\,\ref{table:variancetable}. We set the $x-$axis range to be the same in all sources, to indicate the difference between the spectral coverage of the available data for each source.

The figures show that the variance increases with increasing frequency (except for NGC\,4151). This must be an intrinsic trend. It cannot be due to the increasing contribution of the host galaxy to the observed flux at lower frequencies because we computed the excess variance in physical units, and not normalised by the mean squared. Therefore, the square root of the sample variances indicates the average scatter of the light curve points around the mean in flux units. As such, it is equivalent to the amplitude of the variable part of the observed flux, only. The opposite trend in NGC\,4151 is probably due to the fact that the active nucleus in this source is absorbed by gas and dust in the host galaxy, as we discuss below.

Firstly we fitted the observed variance spectra in the log-log space with a straight line of the form: $\rm log\sigma^{2}_{\nu} = a + b (\rm log\nu - 15)$. During this model fit (and in all model fits in the next sections) we did not consider the SU band points in Mrk\,509 and NGC\,2617 (as variations of the Balmer continuum emitted by the broad line region, BLR,  may contribute to the observed variations in this band), the measurements at wavelengths longer than 6000 \AA\ (for reasons we explain in Sect.\, \ref{sec:7.2}), and the SV \meansxv\ in NGC\,4151 (because the number of segments is small). Dashed lines in Figs.\, \ref{fig:mrk509}, \ref{fig:ngc4151}, \ref{fig:ngc2617}, and \ref{fig:mrk142} show the best-fit lines. The best-fit slopes are: $\rm b = 0.5 \pm 0.4$, $-2.6\pm 0.4$, $0.7\pm 0.8$, and $1.1\pm 0.2$ for Mrk\,509, NGC\,4151, NGC\,2617, and Mrk\,142, respectively. 

The variance spectra are best defined in the case of Mrk\,142 (i.e. the best-fit slope is determined with the smallest error in this case). This is because the Mrk\,142 light curves are long and well sampled and the source is highly variable even on relatively short timescales. This is not surprising, given the fact that Mrk\,142 hosts the smallest BH mass. The variance measurements have larger errors, and the error of the best-fit slope is also large in NGC\,2617. This is probably due to the fact that the \swift\ light curves of NGC\,2617 are not sampled as densely as those of Mrk\,142. The BH mass of NGC\,4151 and NGC\,2617 are comparable and although the duration of the NGC\,4151 light curve segments is smaller than the respective duration in the NGC\,2617 light curves, the NGC\,4151 light curves are much better sampled, hence the variance spectrum is better defined. The \swift\ light curves of Mrk\,509 are also well sampled, but the source is not highly variable on time scales of a few days (as the BH mass is large), so the error of the observed variance measurements (and of the best-fit slope) is relatively large. 

A simple linear model fits the observed variance spectra well, but our objective is to fit the data with the X-ray reverberation model we have developed in the previous sections in order to constrain the physical parameters of the AGNs in the sample (i.e. accretion rate and X-ray corona height).

 \begin{figure}
   \resizebox{\hsize}{!}{\includegraphics{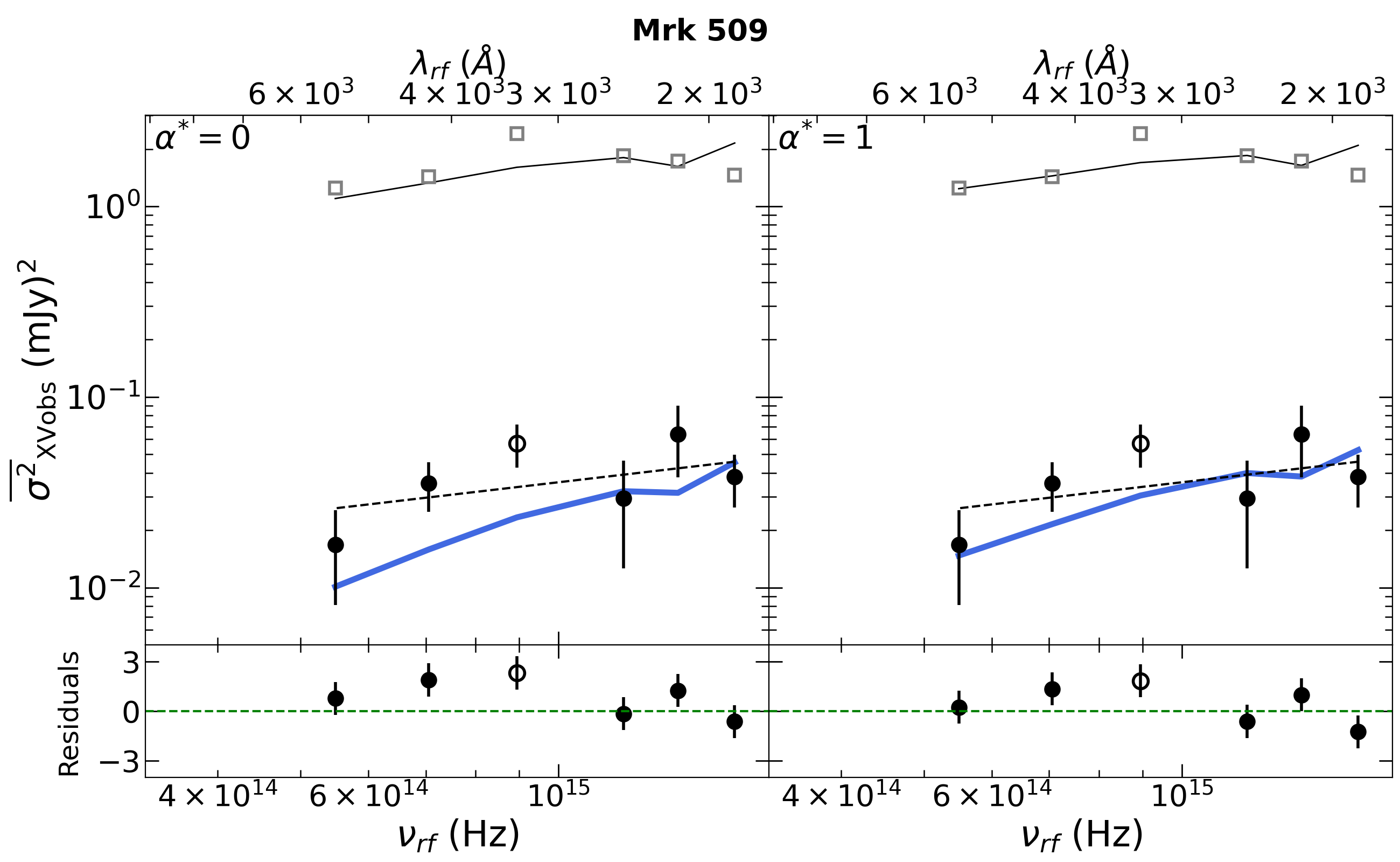}}
      \caption{ Mrk\,509 variance spectrum (i.e. the variance in units of (mJy)$^2$ plotted as a function of photon frequency). Black points show the observed variance measurements calculated from light curve segments of duration, $T_{\rm max}=12$ days (listed in Table\,\ref{table:variancetable}). Open circles indicate data points excluded during the model fits. The black dashed line shows the best linear fit (in log-log space) to the data. The best-fit slope is $\textrm{b} = 0.5 \pm 0.4$.
      The blue lines show the X-ray reverberation best-fit models for $\alpha^{*}=0$ (left panel) and $\alpha^{*}=1$ (right panel). 
      The model's M2 variance is lower than the adjacent W1 and W2 variances because of the $\sim 2200 \AA$ feature in the Milky Way's extinction law as determined by \cite{Cardelli89}. The bottom panels show the best-fit residuals, i.e. the observed variance and best fit model variance over the observed variance error, in the case of the X-ray reverberation best-fit models (blue lines). Grey squares on the top show the measured excess variance calculated using the full light curves, while the solid black line shows the best-fit model prediction for these measurements (see Sect.\,\ref{sec:full} for more details).}
         \label{fig:mrk509}
   \end{figure}

 \begin{figure}
 \sidecaption
   \resizebox{\hsize}{!}{\includegraphics{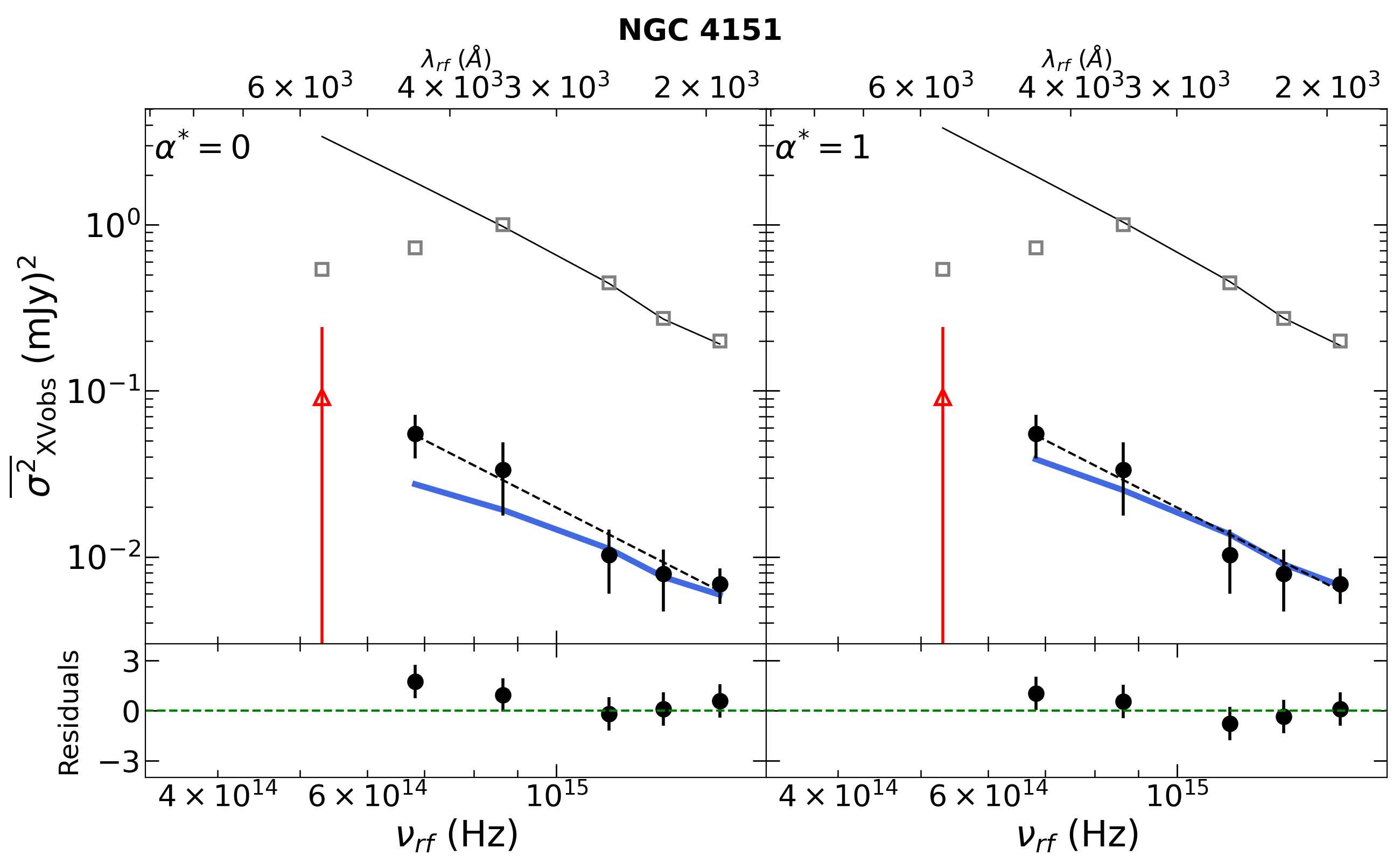}}
      \caption{NGC\,4151 variance spectrum, similar to Fig.\,\ref{fig:mrk509}. The black points are calculated from light curve segments with duration $T_{\rm max}$=3 days, while the red point with $T_{\rm max}$=12 days. The blue line shows the best-fit variance model calculated with $T_{\rm max}$=3. The slope of the best-fit black dashed line is  $\textrm{b} = -2.6 \pm 0.4$. The best-fit variance model for this source includes host galaxy absorption with $E(B-V)=0.35$ (see Sect.\,\ref{sec:notes} for more details). Grey points and solid black line as in Fig.\,\ref{fig:mrk509}.}
         \label{fig:ngc4151}
   \end{figure}

 \begin{figure}
 \sidecaption
   \resizebox{\hsize}{!}{\includegraphics{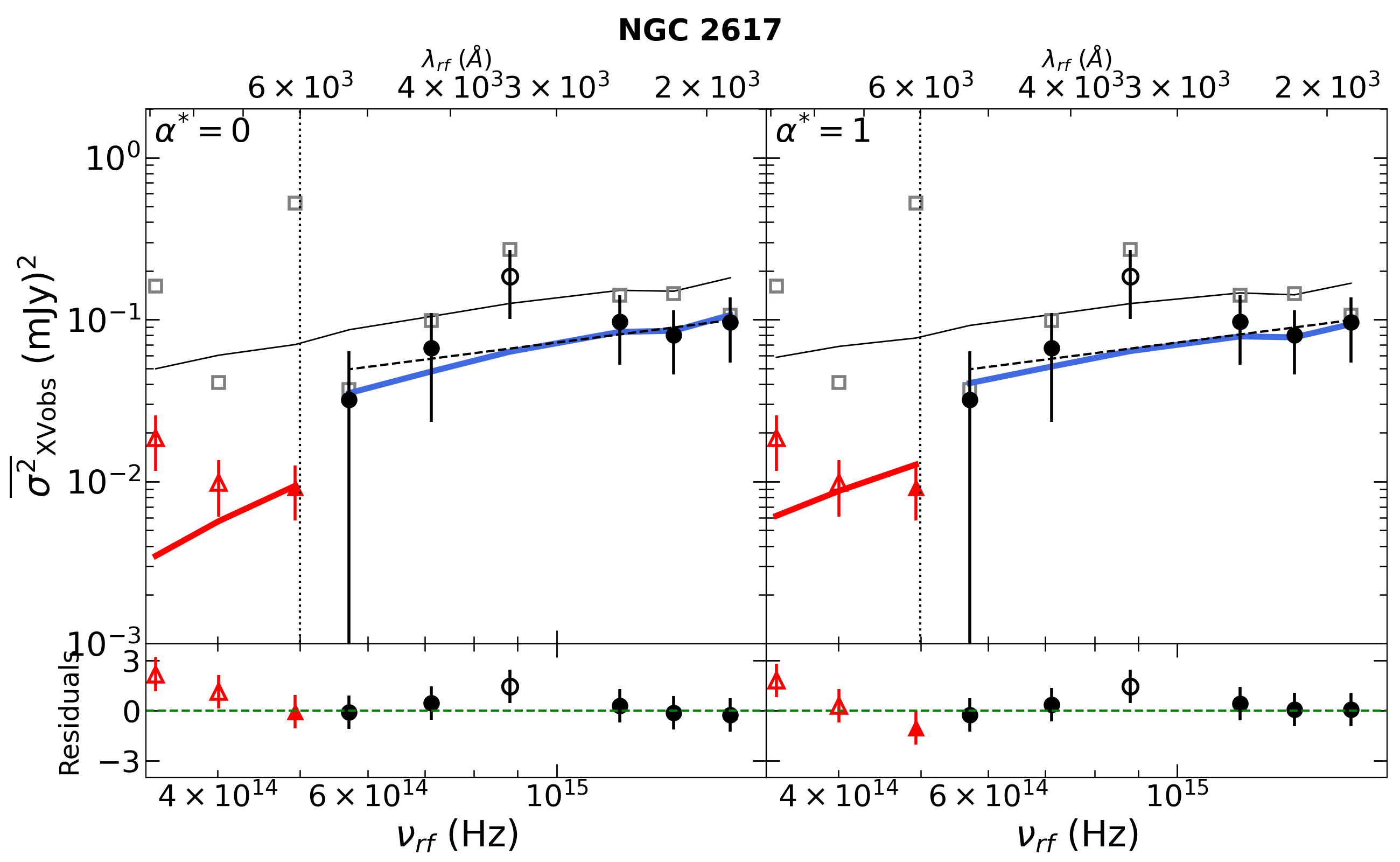}}
      \caption{Same as in Fig.\,\ref{fig:mrk509} for NGC\,2617. 
      We show $\overline{\sigma^2}_{\rm XVobs}$ with $T_{\rm max} = 6$ and 13 days with red triangle and black circle markers, respectively. We show the best-fit model variance for $T_{\rm max} = 6$ and 13 days with red and blue lines, respectively. The M2 variance is lower than the adjacent W1 and W2 variances for the same reason explained in Fig.\,\ref{fig:mrk509}. The slope of the best-fit dashed black line is $\textrm{b} = 0.7 \pm 0.8$. The vertical dashed line indicates $\lambda=6000$\AA. Grey points and solid black line as in Fig.\,\ref{fig:mrk509}.}
         \label{fig:ngc2617}
   \end{figure}

\begin{figure}
\sidecaption
   \resizebox{\hsize}{!}{\includegraphics{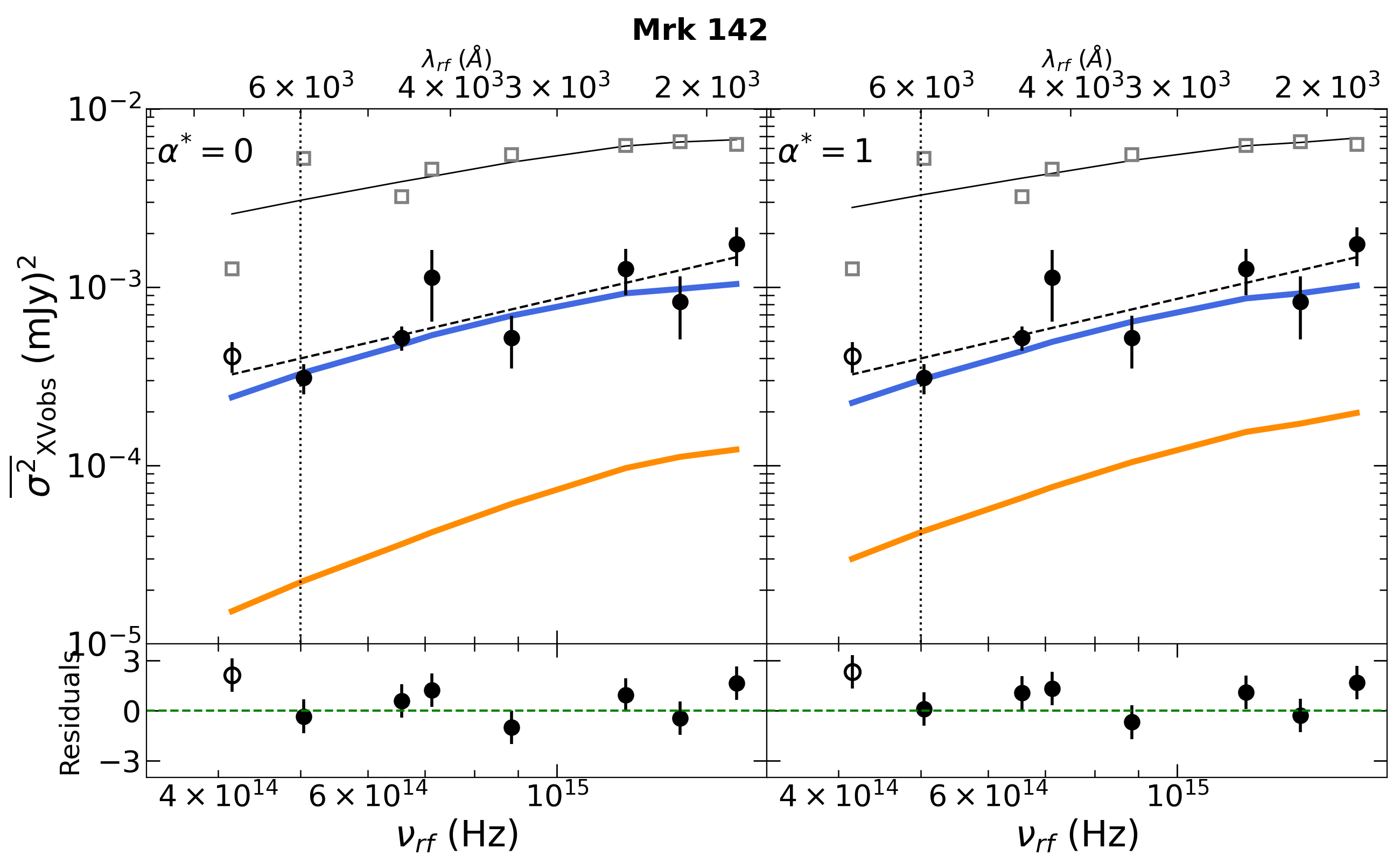}}
      \caption{Same as in Fig.\,\ref{fig:mrk509} for Mrk\,142. The variance measurements have been calculated for $T_{\rm max} = 6$ days. Solid orange lines show the model predictions when assuming $M_{\rm BH} = 2 \times 10^{6}$ and the parameters in K21b (see Sect.  \ref{sec:mrk142}), while the blue lines show the best-fit model for $M_{\rm BH} = 4 \times 10^{6}$ and the parameters listed in Table\,\ref{table:sourcestable} and \ref{table:results}. The slope of the best-fit dashed black line is $\textrm{b} = 1.1 \pm 0.2$. Grey points and solid black line as in Fig.\,\ref{fig:mrk509}.}
         \label{fig:mrk142}
   \end{figure}

\subsection{The model}

We fitted the data plotted in Figs.\,\ref{fig:mrk509}, \ref{fig:ngc4151}, \ref{fig:ngc2617}, and \ref{fig:mrk142} with the model $\sigma^{2}_{\nu, \rm band}$ as defined in Sect.\, \ref{sec:model}. 
We note again that we choose to work in units of (mJy)$^2$ because, in this way, the variance spectra are less steep and it is easier to observe differences between the model and the data visually. The $D_{\rm L}$ are taken from Table\,\ref{table:sourcestable}, and $f_{\rm Gal, \lambda}$ is computed as explained in Sect.\, \ref{sec:modelvar}. We need to consider absorption by the host galaxy only in NGC\,4151 (see Sect.\,\ref{sec:notes}). The disc transfer function, $\Gamma_{\lambda}(\nu)$, depends on 
\spin, \mbh, \medd, $h, L_{\rm Xobs,Edd}$, $\Gamma$, and inclination, $\theta$ (see P22 for details). The BH mass, X-ray luminosity (in Eddington units) and the X-ray spectral photon index are fixed to the values listed in Table\,\ref{table:sourcestable}.  As for the inclination, we assumed $\theta=30^{\circ}$ except for NGC\,4151. \cite{Bentz_2022} studied the geometry and kinematics of the BLR in this object. They found that the BLR is well described by a thick disc with similar opening angle and inclination angle of $\theta\sim 57-58^{\circ}$, suggesting that our line of sight towards the central engine is just above the edge of the BLR. We therefore assumed $\theta=60^{\circ}$ for this object. Regarding the model X-ray power spectrum (i.e. PSD$_{\rm X,Edd}(\nu)$ in Eq.\,(\ref{eq:s2_band_final})), we used Eq.\,(\ref{eq:PSD_XEdd}) with $L_{\rm Xobs,Edd}$ taken from Table\,\ref{table:sourcestable}, and $\overline{\sigma^2}_{\rm HX,NXVobs}$ from Table\,\ref{table:variancetable}. We also computed $\nu_b$ from Eq.\,(\ref{eq:logTb}), using the BH mass and $L_{\rm Bol}$ values listed in Table\,\ref{table:sourcestable}. Consequently, the free parameters of the model are two: the accretion rate, \medd, and the corona height, $h$.

\subsection{Best-fit results}
\label{sec:bestfitres}

We followed a standard $\chi^2$ minimisation procedure to fit the data in the case of \spin=0 and \spin=1. The data displayed with open points in Figs.\,\ref{fig:mrk509}, \ref{fig:ngc4151}, \ref{fig:ngc2617}, and \ref{fig:mrk142} were not considered when fitting the data, for reasons explained in Sect. \ref{sec:observedvariancespectra}. The best-fit $h$ and \medd\ values and their respective 1$\sigma$ confidence region (in brackets), together with their $\chi^2$/degrees of freedom values, are listed in Table\,\ref{table:results} in the third, fourth, and fifth columns respectively). Solid blue and red lines in Figs.\,\ref{fig:mrk509}, \ref{fig:ngc4151}, \ref{fig:ngc2617}, and \ref{fig:mrk142} show the best-fit models over the full frequency range of the data in the plot, irrespective of whether we used the data points during the model fitting. We also show the best model fit residuals in the bottom panels of the same figures. 

The model fits the variance spectra of all sources well. The best-fit accretion rate in many cases is equal to 0.005, which is the smallest value we consider. This is the smallest accretion rate value for which the disc transfer functions of P22 are valid. It seems possible that the model could accommodate even lower \medd\ values while still providing a good fit to the data, albeit with different best-fit $h$ values. This is because \medd\ and $h$ are degenerate, as we explain below. However, our main objective is not to determine the parameter regions that explain the variance spectra on their own but to compare our best-fit results with those presented by K21b to investigate whether the X-ray reverberation hypothesis is sufficient to explain both the time lags and the variance spectra with the same physical parameter values. We can achieve this by comparing the confidence regions of the best-fit parameters as determined by us and by K21b.
    
\begin{center}
   \begin{table}
      \caption[] {
       Best-fit results of this work.}
         \label{table:results}    
\[\arraycolsep=5pt\def\arraystretch{1.2}
         \begin{array}{lcccc}
            \hline
            \noalign{\smallskip}
            \textrm{Source}  & \alpha^{*} & h (R_{g}) & \dot{m}_{\textrm{Edd}}(\%) & \chi^2/\textrm{dof}\\
            \hline
            \noalign{\smallskip}
            \textrm{Mrk } 509 & 0 & 10-22 & 0.5-2.2 & \\
            &   & 13.8\,[10-37]  & 0.5\,[0.5-6] & 5.9/3 \\
            &   1 & 8-18  & 0.5-4 &\\
             &   & 11.4\,[8-37]  & 0.5\,[0.5-2.2] & 4.5/3\\
             \hline
            \textrm{NGC } 4151^\alpha  & 0 & 11-23 & 0.5-1.4 &\\
             &   & 14.5\,[11-93]  & 0.5\,[0.5-19] & 3.1/3\\
            &  1 & 8-22  & 0.5-7 &\\
             &   & 10.8\,[8-63]  & 0.5\,[0.5-44] & 2/3\\
            \hline
            \textrm{NGC } 2617 & 0 & 11-19 & 0.5-1.1 &\\
             &   & 13.5\,[11-22]  & 0.5\,[0.5-2.2] & 5.5/4\\
            &  1 & 11-13  & 6.5-8 \\
             &   & 9\,[6-24]  & 0.85\,[0.5-19]& 5/4\\
            \hline
            \textrm{Mrk } 142 & 0 & 71-100 & 
            18-32  &\\
             &   & 31.6\,[15-100]  & 4\,[0.5-32]& 5.8/5\\
            &  & 3.6\,[2.5-100]  & 52\,[18-56]& 4/8\\
            &  1 & 90-100  &  53-62  &\\
             &   & 50.4\,[17-100]  & 18\,[0.5-62]& 5.7/5\\
            &  & 3.9\,[2.5-100]  & 87\,[53-91] & 4/8\\
            \noalign{\smallskip}
            \hline            
         \end{array}
\]
\tablefoot{The third and the fourth columns list the 1$\sigma$ confidence range of $h$ and \medd\ (in percentage), based on the width of the common area between the two 1$\sigma$ confidence regions plotted in Fig.\,\ref{fig:parameter_space}. Values listed in the line below these results indicate the model variance best-fit $h$ and \medd\ values and the corresponding 1$\sigma$ confidence regions. The last column lists the best-fit $\chi^2$/degrees of freedom (dof) values. For Mrk\,142, we also list our best-fit results to the time lags below the model variance best-fits (see Sect\,\ref{sec:mrk142}).
($^a$) For the variance models of this source, we assumed host galaxy absorption with $E(B-V) = 0.35$.}
\end{table}
\end{center}

 \begin{figure*}
 \sidecaption
    \includegraphics[width=10cm, height=9cm]{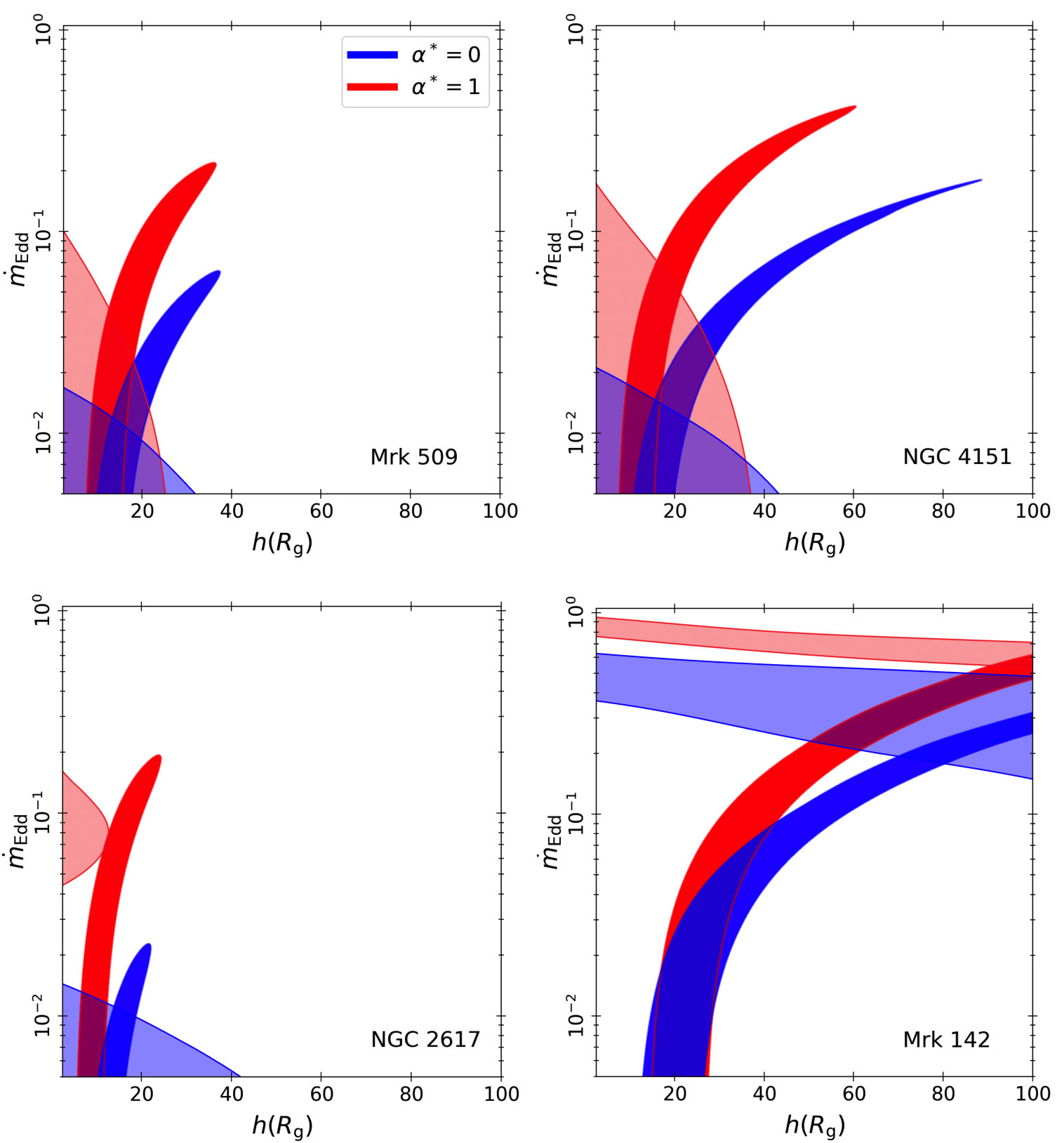}
      \caption{  1$\sigma$ confidence regions of ($h$, \medd) for each source. Blue and red regions indicate the confidence regions for \spin=0 and 1, respectively. Similarly, lighter blue and red shaded areas show the 1$\sigma$ confidence regions as determined by K21b when fitting the time-lags, for \spin=0 and 1, respectively. In the case of Mrk\,142 (lower right panel), the lighter shaded areas indicate the $1\sigma$ ($h$, \medd) confidence regions obtained from our best fit to the time-lags assuming $M_{\rm BH}=4 \times 10^6 M_\odot$ (see Sect.\,\ref{sec:mrk142}).}
         \label{fig:parameter_space}
   \end{figure*}

The combined,$1\sigma$ confidence regions for \medd\, and  $h$ are shown in Fig.\,\ref{fig:parameter_space} with blue-shaded and red-shaded regions (for the $\alpha^*=0$ and $\alpha^*=1$ best-fit models, respectively). The $1\sigma$ confidence regions were determined for a $\Delta \chi^2$ statistic of 2.3 which is equivalent to the 1$\sigma$ confidence region for two interesting parameters (see e.g. \cite{Lampton_76}). The figure shows that the 1$\sigma$ confidence regions are rather large. This could be due to the fact that the sample variances have large errors in some cases, but this is probably not the main reason. For example, the variance spectrum of Mrk\,142 has the smallest errors and the 1$\sigma$ confidence region is still the largest. The large area of the confidence regions can be explained as follows. 

The shape of the confidence regions indicates that the source height and the accretion rate are positively correlated. A larger accretion rate can fit the data equally well if the height of the X-ray source increases. This is because the amplitude of $|\Gamma_{\lambda}(\nu)|^2$ depends on $h$ and \medd\ but in the opposite way: it increases with increasing $h$  but it decreases with increasing \medd\, (see Figs. 8 and 10 in P22, and the discussion in their Sect. 3.2 and 3.3). In fact, $|\Gamma_{\lambda}(\nu)|^2$, depends equally strongly on both parameters and it is for this reason that a rather large range of accretion rate and height values can fit the data equally well. 

K21b fitted the time-lags spectrum of the sources we study using the same X-ray reverberation model. They also fitted the data keeping ($h$, \medd) as free parameters. The lighter red and blue shaded areas in Fig.\,\ref{fig:parameter_space} indicate their 1$\sigma$ confidence region for these parameters (except for Mrk\,142, where the respective confidence region is computed as explained in Sect. \ref{sec:mrk142}). The shape of the combined confidence regions in this case implies an anti-correlation between $h$ and \medd\ (as opposed to the positive correlation implied from the model fits to the variance spectra). This is because the amplitude of the time-lags increases with increasing height and increasing accretion rate as well (see Figs. 17 and 18 in K21a). The range of the parameter values that can fit the time lags well is also large, for the same reason as above: the time lags depend equally strongly on the source height and the accretion rate; therefore, a large range of parameter values can provide an equally good fit to the time-lags.

However, the most important result of our work is that both confidence regions overlap. This result shows that X-ray reverberation can fit both the time lags and the variance spectra of these sources, with the same parameters. Furthermore, we can obtain much stronger constraints on the model parameters when fitting both the observed time lags and the variance spectra. The width of the projection of the intersection between the 1$\sigma$ confidence regions, as defined from the time-lags and the variance spectra model fits, are listed in the third and fourth column in Table\,\ref{table:results}. The values in brackets in the first line below these results list the range of the projection of the 1$\sigma$ confidence region resulting from the variance spectra model fits. For Mrk\,142, we also list the 1$\sigma$ range of the best-fit to the time lags (see Sect. \ref{sec:mrk142}). The size of the common area between the 1$\sigma$ confidence regions should be representative of the 1$\sigma$ confidence region of the model parameters that can fit both the time lags and the variance spectra. This is smaller than the range of the individual 1$\sigma$ confidence regions, which shows that it is necessary to fit both the time lags and the variance spectra to constrain the model parameters as tight as possible in AGNs.

\subsection{Excess variance in the full light curves}
\label{sec:full}

We calculated the excess variance of each source using the full light curves as well. The grey open squares in Figs.\,\ref{fig:mrk509}, \ref{fig:ngc4151}, \ref{fig:ngc2617}, and \ref{fig:mrk142} show the results. In the case of Mrk\,509 and NGC\,4151, we used the full-length \swift\ light curves in all bands, as they are of the same duration. For NGC\,2617, we used the \swift\ light curves from MJD 56640 to 56755 to avoid gaps. In the case of the $r$, $i$, and $z$ bands, we used the light curve from the beginning of the ground-based observations until 115 days after the start, ensuring the same duration for the light curves in all wavebands. Similarly, for Mrk 142, we used light curves from MJD 58485 to 58600 in all wavebands. 

The solid black lines in the same Figures show the model predictions for the excess variance measurements when we consider the full light curves. We used the best-fit results of Table\,\ref{table:results} and we took into account the larger $T_{\rm max}$, and the new \snxvobs\ measurements, computed from the full HX light curve. We note that the model black lines are less steep than the blue ones (in the same figures) since the bias correction term equals unity when the full light curves are used. 

The excess variance measurements are in agreement with the model predictions in the case of NGC\,2617 and they are larger than the model predictions by a factor of $\sim 10, 5.5,$ and 5 in the case of Mrk\,509, NGC\,4151, and Mrk\,142, respectively. This is not surprising, given the fact that the distribution of the variance measurements from the full light curves is not Gaussian and resembles a $\chi^2$ distribution with low effective degrees of freedom \cite[see e.g. the middle and bottom panels in Fig. 3 in][]{Vaughan03}. In addition, the error of the full light curve variance measurements is unknown, as it depends on the value of the true variance of the intrinsic red-noise process \cite[which is of course unknown; see e.g. Section 5.3.3 of][]{Priestley1981}. These facts imply that we cannot properly compare the observed variance and the model predictions. Nevertheless, the overall shape of the model seems to be in good agreement with the observations, at least up to $\sim 6000$\AA. 

For example, the excess of variability amplitude we see in the SU band of Mrk\,509 and NGC\,2617 is present in the variance measurements computed using both the full light curves and the shorter segments. In contrast, no excess of variability amplitude in the same band appears in the variance spectra of NGC\,4151 and Mrk\,142. Furthermore, the long-term variance measurements at wavelengths longer than $\sim 6000$\AA\ scatter erratically around the model predictions in NGC\,2617 and in Mrk\,142 (although with a small amplitude in the latter case). In NGC\,4151 the measurements fall below the model predictions. These variations do not indicate a clear trend and, perhaps, they are due to the fact that the variability amplitude decreases with increasing wavelength in AGNs.

\subsection{Notes on individual objects}
\label{sec:notes}

Mrk 509: 
Figure\,\ref{fig:mrk509} shows that X-ray reverberation can fit well the observed variance spectrum of this source both for \spin=0 and \spin=1, but for a different range of parameters. For example, the \medd\ 1$\sigma$ region is (0.005-0.022) and (0.005-0.04) in the case of \spin=0 and 1, respectively. E19 list an accretion rate estimate of 0.05 for this source. They do not explain how they estimated \medd\ but it is probably computed with the use of a bolometric correction factor. This value is closer to the upper 1$\sigma$ \medd\ limit in the case of $\alpha^*=1$. Although \medd\ estimates based on bolometric correction factors probably have large errors, this result could imply that the BH in Mrk\,509 is rapidly spinning. 

NGC 4151: Initially, we could not find model variance spectra with acceptable $\chi^2$ values for this source. However, it is well known that the central source in NGC\,4151 is absorbed by neutral gas and dust \cite[e.g.][]{Ward87}. To take this into account, we modelled the host galaxy absorption using the reddening law given by Eq.\,(3) in \cite{Czerny_04}, and we considered $E(B-V)$ values in the range of 0-1, with a step of 0.05. For each $E(B-V)$ value we computed $f^2_{\rm host, \lambda}$ (as explained in Sect.\,\ref{sec:modelvar}) and we applied it to the estimation of the model variance spectra (as shown in Eq.\,\ref{eq:s2_band_final}). The best agreement with the data was achieved for $E(B-V)=0.35$. Figure\,\ref{fig:ngc4151} shows that when accounting for intrinsic absorption the model can describe the data well. 

An $E(B-V)=0.35$ value translates to a hydrogen column density of $N(H)\approx 2.9 \times 10^{21}$ cm$^{-2}$ \citep{Liszt_2021}. This is a rather small column density for the neutral absorber in NGC\,4151 \cite[e.g.][]{Beuchert2017,Lubinski16}. However, $N(H)$ is variable in this source. K21b fitted the mean X-ray spectrum of the source during the \swift\ monitoring campaign assuming an ionised absorber only. This implies that the column density of the neutral absorber may be relatively small during the \swift\ observations of NGC\,4151 in 2016.

NGC 2617: Figure\,\ref{fig:ngc2617} shows the results for NGC\,2617. The X-ray reverberation model fits the variance spectrum well. The number of the light curve segments we used to compute the excess variance is the smallest among the sources in our sample (in all bands). 
Consequently, the errors may not be Gaussian and the results from the model fitting when using $\chi^2$ statistics may not be accurate in this case. Figure\,\ref{fig:parameter_space} shows that the 1$\sigma$ confidence region of the model parameters from the fits to the time-lags and the variance spectra barely overlap for \spin=1. This could be due to the fact that the error of the variance measurements may not be Gaussian (the range of the 1$\sigma$ confidence region depends directly on the assumption of the errors being Gaussian). In any case, this is not a reason to claim that a maximally spinning BH is excluded for this source, as we consider the common area of the 1$\sigma$ confidence regions only, which is not highly significant. On the other hand, F18 report an accretion rate estimate of 0.01 for this source. This is within the range of the common area between the $1 \sigma$ regions that we report for \spin=0 in Table\,\ref{table:results}. This result could indicate a non-rotating BH in this AGN.

\subsection{The case of Mrk 142}
\label{sec:mrk142}

 \begin{figure}
   \centering
    \includegraphics[width=8cm, height=6cm]{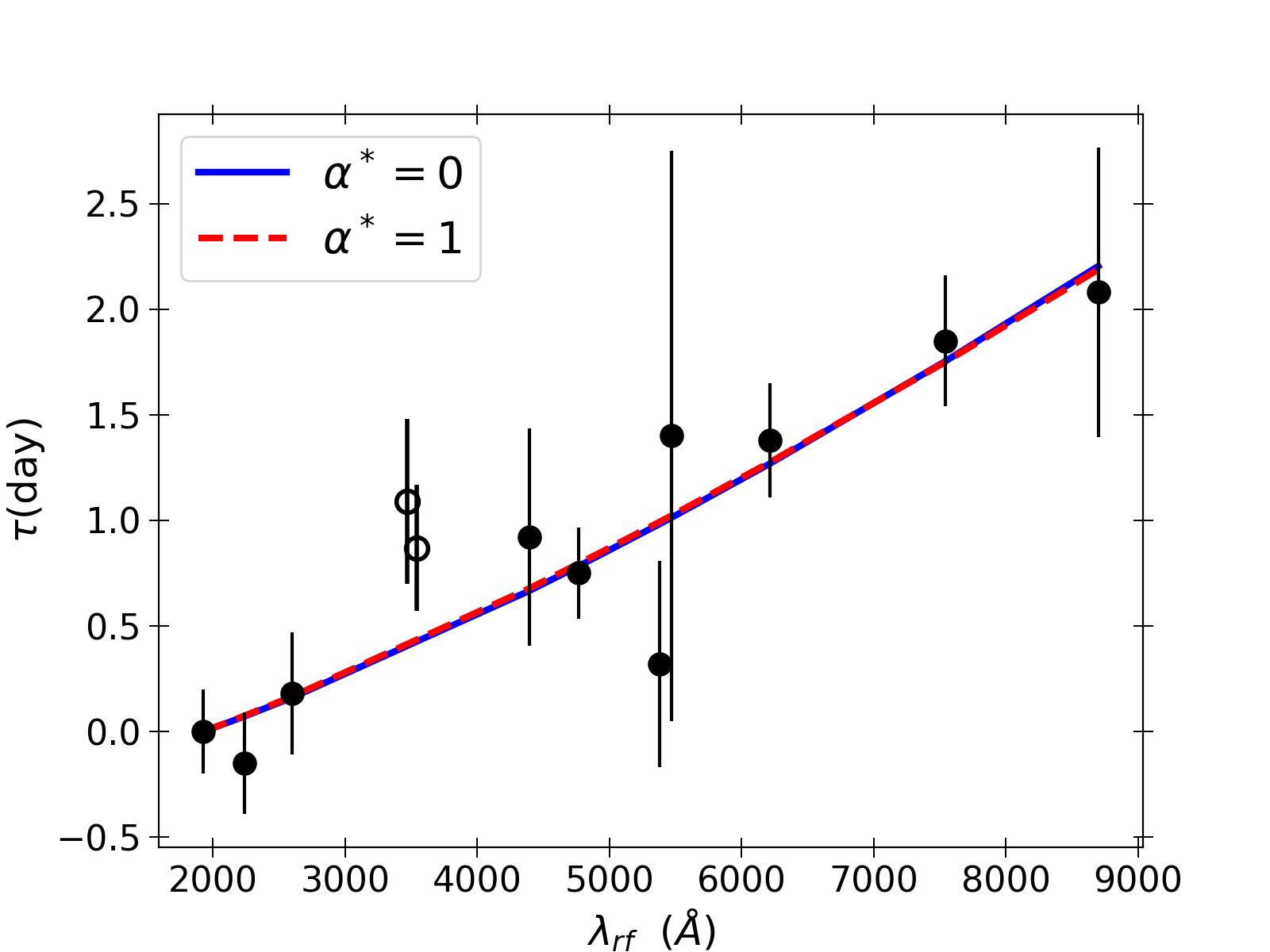}
      \caption{Time-lag spectrum of Mrk\,142 (data taken from \cite{Cackett2020}). The blue solid and red dashed lines show the best-fit model time-lags for $\alpha^{*}=0, 1$, while the parameter values are listed in Table\,\ref{table:sourcestable}.}
         \label{fig:timelags}
   \end{figure}

Black circles in Fig.\,\ref{fig:mrk142} show the measured variance spectrum of Mrk\,142. The orange-coloured solid lines show the model variance spectra calculated using the model parameters that K21b assumed for this source, namely: $M_{\rm BH}=2 \times 10^{6}M_\odot$, $L_{\rm Bol}=2.1 \times 10^{44}$erg/s, $L_{\rm Xobs, Edd}=0.0246$, and their best-fit $h$ and \medd\ values, which are listed in their Table\,3 (i.e. $h=51.1$ and \medd=0.64 for \spin=0, and $h=50.9$ and \medd=0.88 for \spin=1). The model predictions are significantly smaller than the observed variance spectrum. This disagreement could be a problem for the X-ray reverberation model, so we investigated possible reasons for this.

To increase the model variability amplitude we need to increase the normalisation of $|\Gamma_{\lambda}(\nu)|^2$. Figures\,7--12 in P22 show how the transfer function amplitude depends on the various model parameters. The amplitude can increase with decreasing inclination (although the difference between the amplitude of $|\Gamma_{\lambda}(\nu)|^2$ with $\theta=30^{\circ}$ and $\theta=0^{\circ}$ is small), increasing $\Gamma$ (however, with $\Gamma$  already being very steep for this source), or increasing height. The amplitude of $|\Gamma(\nu)|^2$ will change by a factor of 2-3 if we increase the height from $\sim 50$ to 100\rg, which is the largest height that K21a and P22 considered; however, even this increase is not enough to explain the discrepancy between the model and the observed variance spectra in Fig.\,
\ref{fig:mrk142}. So, we also considered the possibility that the BH mass is slightly larger than the value assumed by K21b. It is commonly assumed that the BH mass estimates in AGNs are affected by systematic errors of the order of $\sim 0.3$ dex (i.e. by a factor of 2 or so, on average). We therefore investigated the possibility that \mbh=$4\times10^6$ M$_{\odot}$ in Mrk\,142 (i.e. two times larger than the value assumed by K21b). We note that this value is consistent almost within 1$\sigma$ with the BH mass measurements of \cite{Li_2018} and \cite{Dupu14}.

We assumed a bolometric luminosity of $L_{\rm Bol} = 1.85 \times 10^{44}$erg/s, as computed by \cite{Cackett2020} when using the relation of $L_{\rm Bol}=9\lambda L_{\lambda} (5100 \AA)$ of \cite{Kaspi2000}. We also considered the 2--10 keV X-ray flux measurement of K21b, but since we assumed a larger \mbh, $L_{\rm Xobs, Edd}$ is now smaller. The chosen values for \mbh, $L_{\rm bol}$, and $L_{\rm Xobs, Edd}$ are listed in Table\,\ref{table:sourcestable}. For these parameters, the X-ray reverberation model can fit the variance spectrum of the source very well (best-fit results are listed in Sect. \ref{sec:bestfitres}). The best-fit models are shown by the blue solid lines in Figure\,\ref{fig:mrk142} and the $1\sigma$ confidence regions for $h$ and \medd\ are shown in the lower-right panel of Figure\,\ref{fig:parameter_space}. We note that the large difference between the blue and orange models in Fig.\,\ref{fig:mrk142} is not only due to the change in mass. According to the results of \cite{Panagiotou22}, the amplitude of the transfer function (squared) increases with increasing BH mass as $\sim M^3_{\rm BH}$. Therefore, the increase in BH mass implies that the blue line model should be about 8 times higher than the orange one. However, the increase in the BH mass implies a smaller accretion rate (for a fixed bolometric luminosity). This resulting decrease in the best-fit \medd\ explains the remaining difference between the two model lines. 

Since the \mbh\ and $L_{\rm Xobs,Edd}$ values that K21b assumed when fitting the Mrk\,142 time-lags are different to the values we assumed when fitting the variance spectrum, we needed to investigate whether the X-ray reverberation model was also able to fit the observed time-lags of Mrk\,142 when using the values listed in Table\,\ref{table:sourcestable}.  To this end, we used the time lags model defined by Eqs. (1)-(9) of K21b and we fit the time lags spectrum by keeping the BH mass and the X-ray luminosity fixed to the values listed in Table\,\ref{table:sourcestable}. Figure\,\ref{fig:timelags} shows the observed time-lags with black points (data taken from C20) and the best-fit models with blue solid and red dashed lines for spin-zero and one, respectively. We did not consider the time-lags around $\sim 3000$\AA\ when fitting the data (i.e. points plotted with grey open circles in Fig.\,\ref{fig:timelags}) for the reasons explained above. The $1\sigma$ confidence region for ($h$, \medd) from the model fits to the time-lags is shown with the lighter blue and red shaded areas in the lower right panel of Fig.\,\ref{fig:parameter_space}. The respective best-fit results, the 1$\sigma$ confidence regions, and the $\chi^2$/dof values are listed in Table\,\ref{table:results}. If we consider the \mbh\ and $L_{\rm Bol}$ values listed in Table\,\ref{table:sourcestable}, then \medd\ should be $\sim 0.4$ for this source. This is higher than the upper limit of the common 1$\sigma$ confidence region that we find for \spin=0. This could be an indication that just like Mrk\,509, the BH in this source may be rotating; however, it is not rotating at a maximum rate, as the lower limit of the common 1$\sigma$ region is larger than 0.4 for \spin=1.


\section{Summary and discussion}
\label{sec:discussion}

We studied the optical/UV variance spectra (i.e. the excess variance as a function of photon frequency) in four nearby Seyferts, namely: Mrk\,509, NGC\,4151, NGC\,2617, and Mrk\,142, using data from recent, long, multi-wavelength campaigns with \swift\ and ground-based telescopes (in two sources). We measured the excess variance in various spectral bands and compared it with the model variance predicted by the X-ray reverberation model. To compute the model variance, we employed the accretion disc transfer functions of \cite{Panagiotou22} and assumed a bending power law for the X-ray power spectrum. 

We found that the X-ray thermal reverberation model can fit well the variance spectra of the sources in the sample. However, the most important result of our work is that X-ray reverberation can explain both the variance spectra and the time-lags in these AGNs for the same physical parameters. To the best of our knowledge, this is the first time that the X-ray reverberation model has been shown to simultaneously fit well, with respect to both the time lags and variability spectra of an AGN. 

\subsection{On the X-ray reverberation model fits to the variance spectra and the time-lags}

It has been stated many times in the past that X-ray reverberation predicts time lags that are smaller than the observed ones by a factor of approximately 2 to 3. This was first noticed by \cite{Edelson2015} and \cite{Fausnaugh16}, who studied the X-ray optical/UV times lags in NGC\,5548. We have repeatedly shown in the past that X-ray reverberation can fit  the observed time lags in many sources well. In fact, it can do so, for a large range of X-ray height and accretion rate values. In this study, we show that X-ray reverberation can also accurately account for the observed UV/optical variability amplitude in the four AGNs we studied (see also P22 for the case of NGC\,5548). Furthermore, akin to the findings of K21b and Kammoun et al. (2023), we find that the X-ray reverberation model not only fits the variance spectra well but also does so across a broad range of $h$ and \medd\ values (see Fig.\, \ref{fig:parameter_space}). 

K21b have discussed in detail what are the differences between the previous, analytical, X-ray reverberation time-lag models, and the X-ray reverberation models that were developed by K21a (and P22). A major difference that can significantly affect  the model predictions for the time lags and variance spectra is that previous models assumed that the local ratio of external heating (due to X-ray illumination) to internal heating (due to accretion) is constant with the radius measurement. The observations show that this is not the case. One reason for the ability of the K21b and P22 models to fit well the observations is due to the fact that the model computes the external to internal heating at each radius, for any combination of $h$ and \medd. This is important, as this ratio depends strongly on both of these physical parameters. As the accretion rate increases, the temperature of the disc rises as well. Consequently, for a constant X-ray luminosity and $h$, a smaller fraction of the observed emission is due to X-ray reprocessing (in each energy band); hence, the variability amplitude drops. On the other hand, as the corona height increases, the disc’s solid angle, as seen by the X-ray source, increases as well. As a result, for a constant \medd\ (and X-ray luminosity), the amount of X-ray emission being reprocessed in the disc is larger and the variability amplitude increases (in all bands, but not in the same way). It is exactly for these reasons that X-ray reverberation can fit the data for a large range of parameters (as we also explain in Sect. \ref{sec:bestfitres}).

The main result of our work is that the X-ray reverberation model can fit well both the time lags and the variability amplitude of the AGNs we study here, for a reasonable range of parameters. However, we cannot be entirely certain about the range of the allowed parameter values, as we have used a particular X-ray reverberation model. Our modelling is based on the transfer functions of P22 which, in turn, were based on the disc response functions of K21b. The latter were computed only for two spin values, using a colour correction factor of $f_{\rm col}=2.4$\footnote{This means that each disc element is assumed to emit as a modified blackbody, that is a blackbody normalised by a factor $1/f^4_{\rm col}$ and with temperature boosted by a factor $f_{\rm col}$, thus keeping the same total emitted flux.} and assuming that the X-ray source is powered by an unknown source that is not related to the power released by the accretion process. As a result, the best-fit parameters reported here are only valid for the above assumptions as well.  \cite{Kammoun23}  recently released a new code which can compute the disc transfer functions for any spin value and colour correction factor for when the X-ray source is powered by the accretion process or from a different source of power. We plan to use the new code for the study of the variability amplitude of AGNs in the future. We believe that our main result that X-ray reverberation can account for the variability amplitude and the time lags in these sources will remain valid, although the range of allowed parameters may change. For example, \cite{Kammoun212} found that the time lags of NGC\,4593, and NGC\,5548 can be aptly fitted with \spin=0 and 1 (assuming $f_{\rm col}$=2.4), while \cite{Kammoun23} found that the time-lags of the same sources can be well fitted by also assuming \spin=0.7 and  $f_{\rm col}$=1.7. According to the \cite{Kammoun212} results, the 1$\sigma$ corona heights range in NGC\,5548 and NGC\,4593 is 23-59/33-78 \rg\ and 2.5-25/2.5-23 \rg\ (for \spin=0/1, respectively), while \cite{Kammoun23} found heights greater than 64 \rg\ (for NGC\,5548) and a range of 5.6-43 \rg\ for NGC\,4593 for their solution of \spin=0,0.7 and $f_{\rm col}$=1.7 (when 50\%\ of the accretion power is transferred to the corona). \cite{K24} showed that the X-ray reverberation model can also explain well the
time-resolved SEDs of NGC\,5548 for spin=0 and $f_{\rm col}$=1.7. Although the range of the physical parameters may not be exactly equal to the values listed in Table \ref{table:results} when we consider a different $f_{\rm col}$ and/or spin, the main result will remain the same: X-ray reverberation can explain both time-lags and the variance spectra in these sources. 

As we argued above, the X-ray reverberation model can fit the time lags for various parameter values. In order to constrain the range of parameters that can explain the observational properties of AGNs, we should also consider additional observables. In a future publication, we plan to study time-lags, variance spectra and the mean, broadband SED, determined by the same data which are also used to compute time-lags and variance spectra. 

\subsection{Considering whether X-rays drive all the observed fast, optical/UV variations in AGNs}
\label{sec:7.2}

An excess of variability amplitude appears around $\sim 3000$\AA\ in Mrk\,509 and, perhaps, in NGC\,2617 as well, but it is not statistically significant. An increase in the time-lags spectra\footnote{A small excess at around 3000\AA\  appears in the time-lags spectra of the sources in our sample, except for NGC\,4151 (see Fig.\,3 in K21b). K21b did not take into account the time-lags at this wavelength when fitting the time-lags spectra, although the discrepancy between the measured time-lags at this wavelength. The best-fit model was mildly significant only in the case of Mrk\,509.} in the same energy band is attributed to diffuse emission from gas in the much larger BLR (in terms of  size), which should contribute to the observed flux between 2000-4000 \AA\ \cite[e.g.][]{Netzer22,Korista19}. If such a component is present, it should increase the variance in these wavelengths by an amount that should be representative of the percentage of this component's flux over the total flux in the band. Our results show that such a component is not significant in the variance spectra of the four sources we study. 

Figures\,\ref{fig:ngc2617} and \ref{fig:mrk142} show that the measured variability amplitude may be larger than the model predictions approximately after $\lambda = 6000\AA$ (marked with the vertical dashed lines) in NGC\,2617 and Mrk\,142. 
This could be an indication that X-ray reverberation cannot account for all the variable flux at large wavelengths. However, as K21a argued, their response functions at long wavelengths are truncated because they were computed for a disc with $R_{\rm out}=10^4 R_g$. For example, Fig.\,9 in K21a shows that the response functions at $\sim 10000$ \AA\ for a $10^7$ M$_{\odot}$ black hole are truncated, even when \medd=0.01. Such truncation effects could be more significant in NGC\,2617 at long wavelengths (and even more significant if the accretion rate is larger than 0.01 of the Eddington ratio) and they could explain the slight disagreement between the model and the data variances at long wavelengths. An increased variability amplitude at longer wavelengths could also indicate that the disc is not entirely flat. If the disc height to radius ratio is constant, then we expect the disc height to increase at large radii. In this case, X-ray absorption will increase; hence, the amplitude of the variable flux component will be larger than that predicted by the P22 models, which assumed a flat disc at all radii. This effect should probably be more pronounced at longer wavelengths as the disc height will be quite large at large radii. It is for these reasons that we did not include $\overline{\sigma^2}_{\rm XVobs}$ measurements at wavelengths longer than 6000\AA\ in the model fits. However, we stress that the disagreement between the best-fit model and the variance measurements at long wavelengths is not significant. 

We therefore conclude that all of the observed optical/UV variability in these AGNs, on the sampled time scales (i.e. a few days up to a week) could be due to X-ray reverberation. It is difficult to explain such fast variability with intrinsic disc variations, as the accretion disc characteristic time scales in AGNs are much longer. For example, the viscous and thermal time-scale at a distance of 100 Schwartzchild radii (so that any variation at this distance will affect both the UV and the optical emitting areas of the disc) is of the order of 800 and 80 days, respectively, for a black hole of $10^7$ M$_{\odot}$ mass. Our result of an intrinsically non-variable disc is in agreement with results from variability studies of Galactic X-ray black-hole binaries. The accretion disc in their soft state appears to be intrinsically constant \cite[e.g.][]{Churazov01}, as is the case for the AGNs we consider here. However, the accretion disc may be intrinsically variable on longer time scales. A variability analysis of light curves on longer time scales will allow us to address this possibility (in a future publication).

\subsection{Height of the X-ray source and the BH spin in AGNs}

Our results can put some interesting constraints on the height of the X-ray source and the BH and spin in AGNs. In agreement with K21b, our results indicate that the BH mass may be rotating in Mrk\,509 and Mrk\,142. This is because the accretion rate estimates for these objects (e.g.\, 0.05 for Mrk\,509, from E19, and 0.4 for Mrk\,142, computed as explained in Sect. \ref{sec:mrk142}) are larger than the upper limit of the 1$\sigma$ confidence region that is listed in Table\,\ref{table:results} (for \spin=0). In fact, we computed the common 3$\sigma$ confidence regions for \medd\ and height, and we found that these accretion rate estimates are not consistent with the hypothesis of \spin=0\ at the 3$\sigma$ level. On the other hand, the accretion rate estimate of 0.01 for NGC\,2617 is within the $1\sigma$ confidence region if we assume \spin=0. In this case, we can exclude a maximally spinning BH at the 2.5$\sigma$ level.


Our results suggest that the height of the X-ray source is $\sim 10-20 R_g$ in Mrk\,509, NGC\,2617, and NGC\,4151. These are relatively low heights, close to the lowest height limit below which it would be difficult to detect the UV/optical variations due to X-ray reverberation in these objects. Indeed, as K21a showed (see their Fig.\,26), the fraction of the variable flux (due to X-ray reverberation) over the total UV/optical flux decreases significantly with decreasing height below $\sim 20 R_g$.  The X-ray source height in Mrk\,142 turns out to be larger than 70$R_g$. This is the source with the smallest BH mass and the largest accretion rate among the sources in the sample. However, our sample is too small for us to investigate whether the source height depends either on BH mass or accretion rate. We need a larger sample to start investigating these trends. 

Our work highlights the importance of modelling the optical/UV variability in the long, densely sampled light curves from the recent, multi-wavelength monitoring campaigns of many AGNs. Most of the analysis so far has focussed on the study of the correlation between the X-ray and the optical/UV light curves. However, the study of either the variance (as we do in this work) and/or the power spectrum as a function of wavelength can also help constrain theoretical models. We believe that the study of both the time lags and of the variability amplitude as a function of wavelength, simultaneously, is the strongest test of any model that tries to explain the optical/UV variability in AGNs. We show that the X-ray reverberation can explain both the time lags and the variance spectra for a reasonable set of model parameter values. We plan to study more objects in the future, with the aim of determining the physical parameters of the inner engine in these objects. 

\begin{acknowledgements}
We thank the referee for their comments which helped us improve the manuscript. M.P. acknowledges support from the European Research
Council (ERC) under the European Union Horizon 2020 research and innovation program under grant agreement No 771282.
\end{acknowledgements}

\begin{center}
    \textbf{ORCID iDs}
\end{center}
\noindent
Marios Papoutsis \orcidlink{\orcidauthorA} \href{https://orcid.org/0009-0009-8988-0537}{https://orcid.org/0009-0009-8988-0537}\\
Iossif Papadakis \orcidlink{\orcidauthorB} \href{https://orcid.org/0000-0001-6264-140X}{https://orcid.org/0000-0001-6264-140X}\\
Michal Dovčiak \orcidlink{\orcidauthorC} \href{https://orcid.org/0000-0003-0079-1239}{https://orcid.org/0000-0003-0079-1239}\\
Elias Kammoun \orcidlink{\orcidauthorD} \href{https://orcid.org/0000-0002-0273-218X}{https://orcid.org/0000-0002-0273-218X}

%
%
\bibpunct{(}{)}{;}{a}{}{,} 
\bibliographystyle{aa}
\bibliography{main}

\begin{appendix}

\section{The light curves and variance measurements of the sources}
\label{app:A}

 \begin{figure}[b]
 \resizebox{\hsize}{!}{\includegraphics{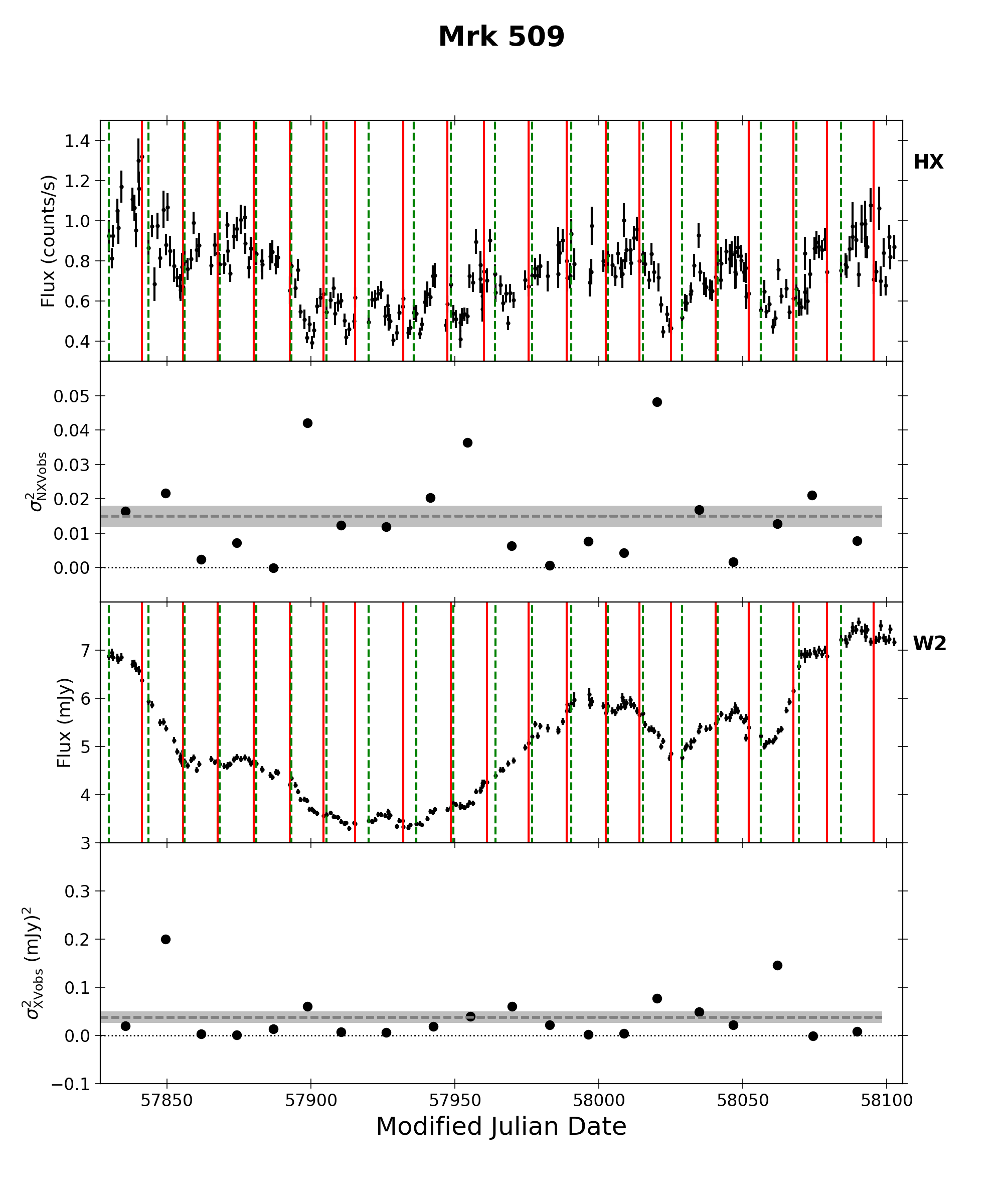}}
      \caption{ X-ray and \swift\ W2 light curves of Mrk\,509 (data taken from E19). The green dashed and red solid lines indicate the starting and ending points of the light curve segments that we used to compute \snxvobs\  and \sxvobs. The segments we used to compute the excess variance in the other optical bands are very similar to the ones indicated in the panel which shows the W2 light curve. The black points in the bottom panels of each plot show the variance measurements of each segment. The grey dashed lines and the grey area show the mean variance and its error, respectively. We draw black dotted lines at \snxvobs\= 0 and \sxvobs\= 0.}
         \label{fig:mrk509_lightcurve}
   \end{figure}
   
 \begin{figure*}
 \sidecaption
   \centering
    \includegraphics[width=12cm]{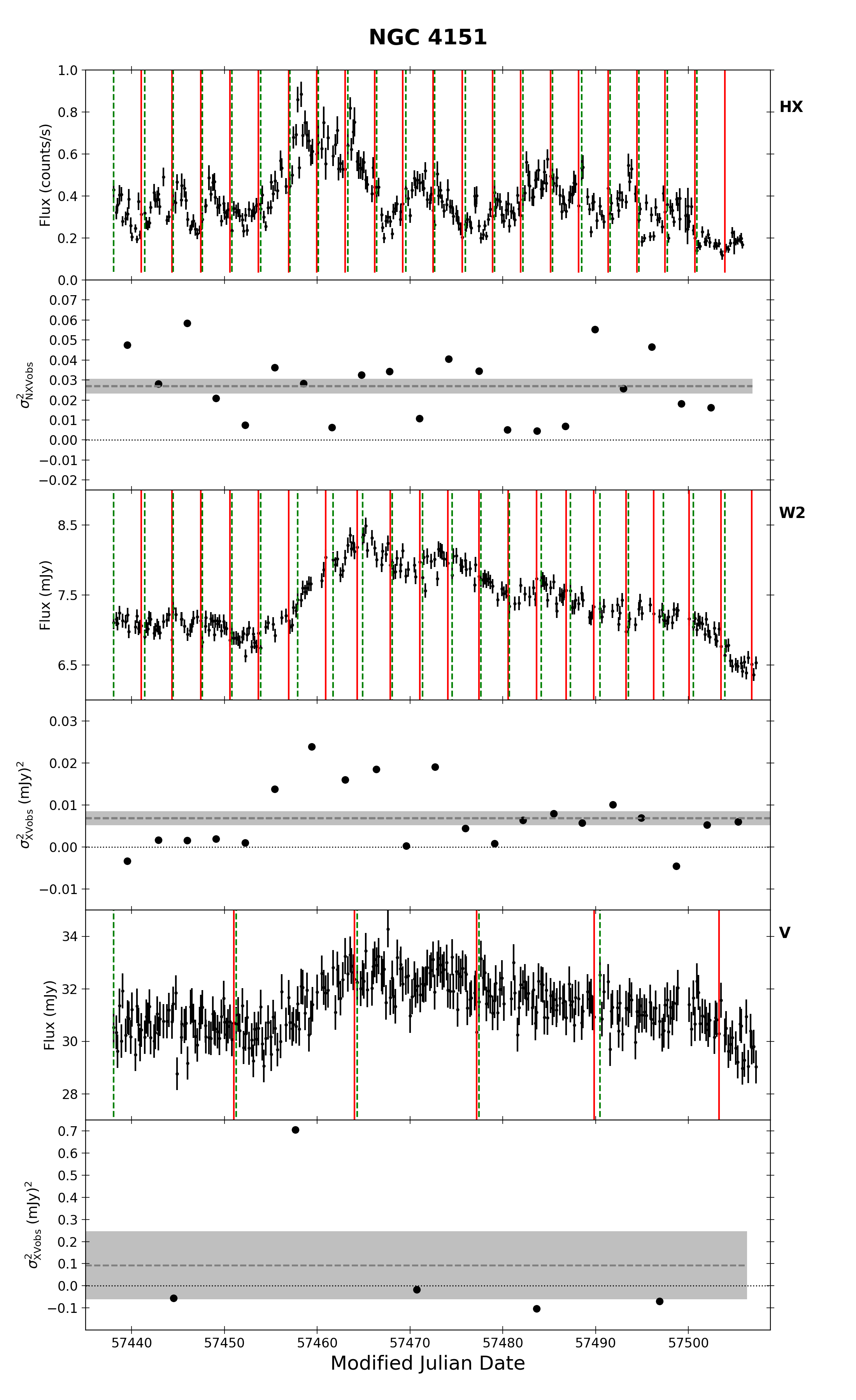}
      \caption{ Same as in Fig.\,\ref{fig:mrk509_lightcurve} for NGC\,4151 (data are taken from E17). In this case, we plot the \swift\ V band light curve as well, as the width of the segments in this light curve is different than the width of the segments in the \swift\ W2 and the other band light curves. }
         \label{fig:ngc4151_lightcurve}
   \end{figure*}

 \begin{figure*}
 \sidecaption
    \includegraphics[width=12cm]{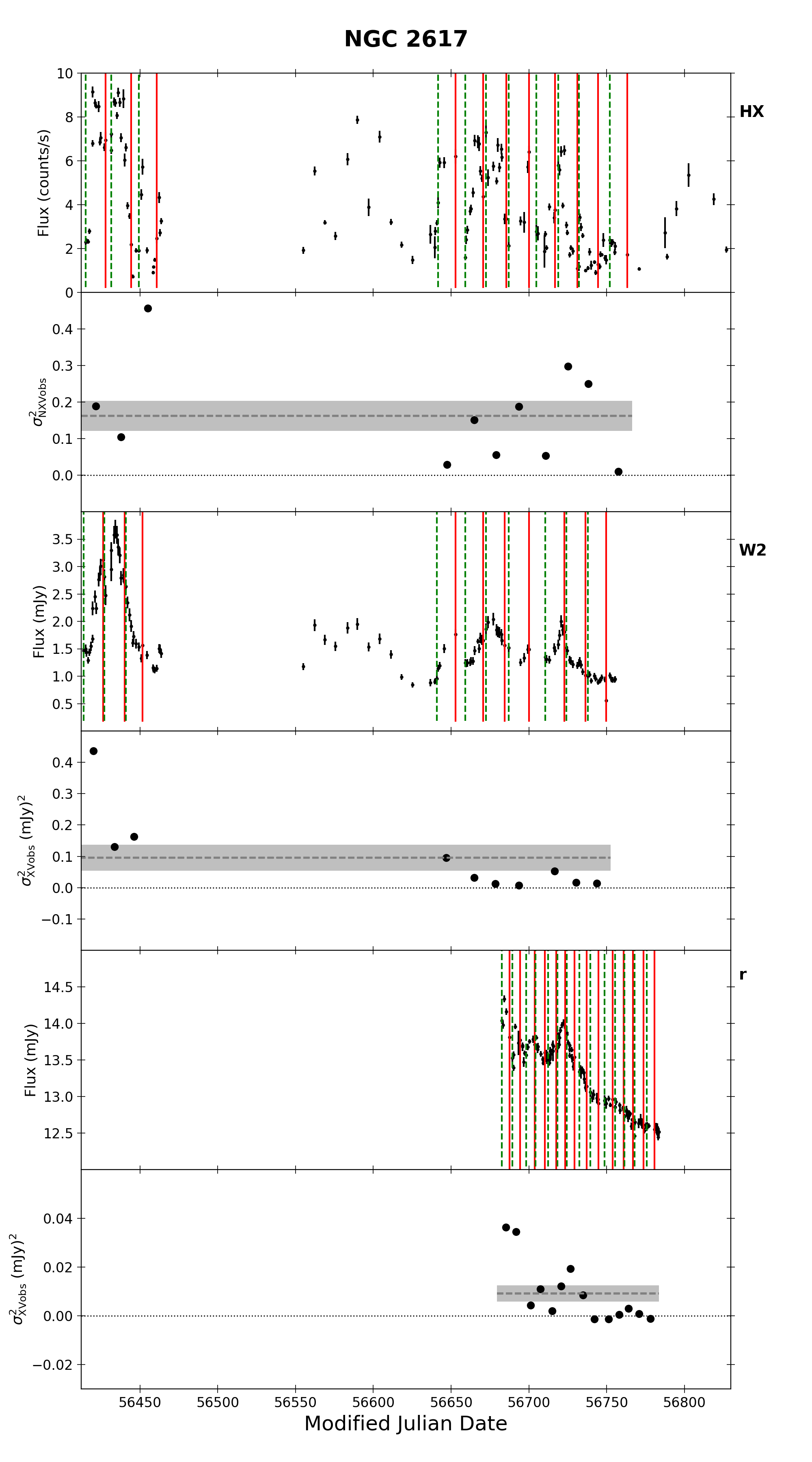}
      \caption{ Same as in Fig.\,\ref{fig:mrk509_lightcurve} for NGC\,2617 (data taken from F18). To extend the spectral coverage of the variance spectra, we also considered light curves in the $r, i,$ and $z-$bands for this source (data from the same source). For this reason, we also plot the $r-$band light curve in this figure. The segments we used to compute the excess variance in the $i$ and $z-$band light curves are very similar to the ones indicated in the $r-$band light curve.}
         \label{fig:ngc2617_lightcurve}
   \end{figure*}

 \begin{figure*}
 \sidecaption
    \includegraphics[width=12cm]{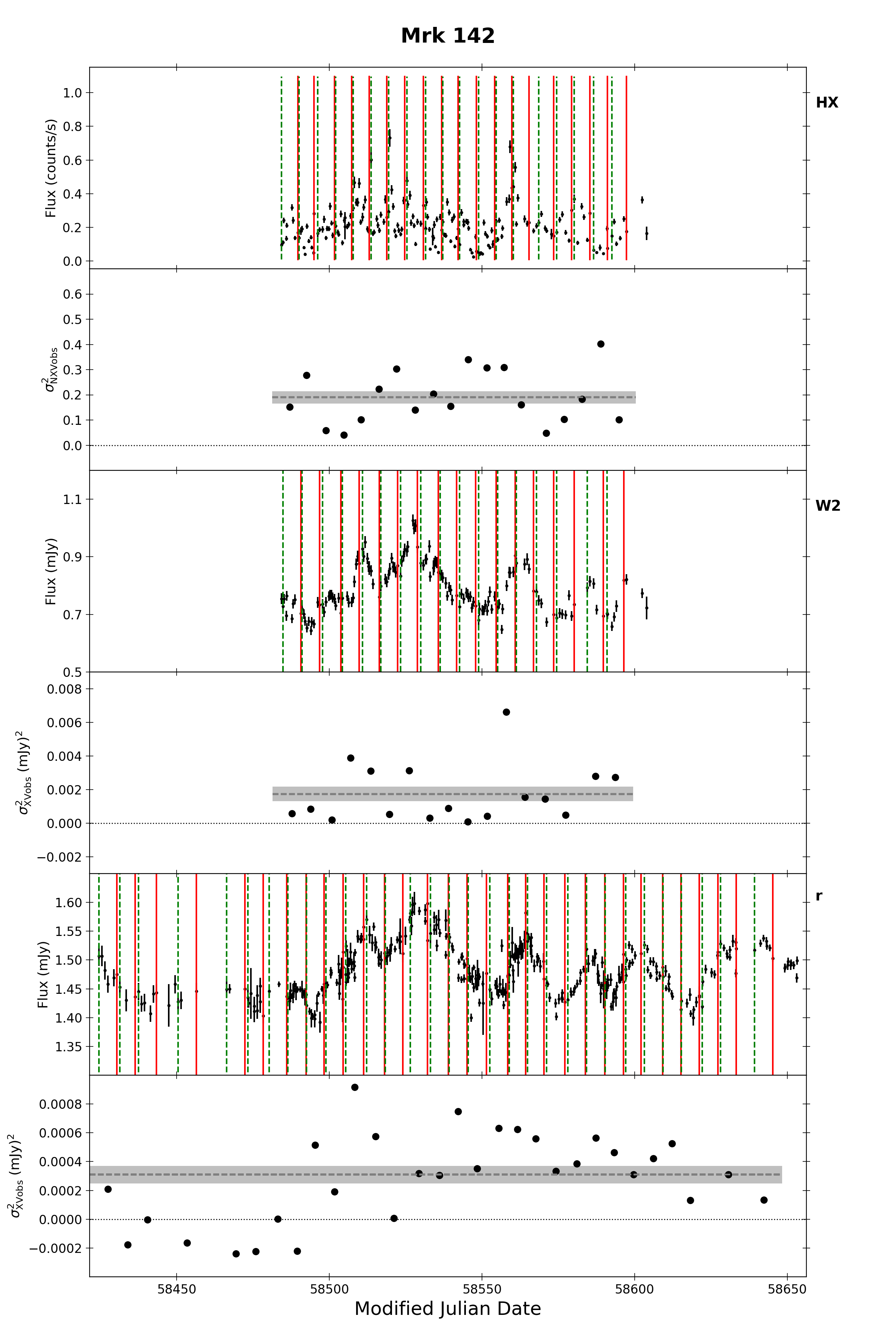}
      \caption{ Same as in Fig.\,\ref{fig:ngc2617_lightcurve} for Mrk\,142 (data taken from C20). To increase the spectra coverage and the number of variance data within the spectral width of the \swift\ filters we also used light curves from ground-based telescopes for this source. These light curves are well sampled and significantly longer than the \swift\ light curves. We plot the $r-$band light curve in this figure; the segments we used to compute the excess variance in the $u, g. i$ and $z-$band light curves (data taken from C20 as well) are very similar to the ones indicated in the $r-$band light curve. }
        \label{fig:mrk142_lightcurve}
   \end{figure*}

\end{appendix}
\end{document}